\documentclass{emulateapj}
\usepackage{epsfig}
\usepackage{graphicx}
\usepackage{color}
\usepackage[normalem]{ulem}

\newcommand{\be}{\begin{eqnarray}}
\newcommand{\ee}{\end{eqnarray}}
\newcommand{\beq}{\begin{equation}}
\newcommand{\eeq}{\end{equation}}
\renewcommand{\vec}[1]{\mbox{\boldmath $\displaystyle #1$}}
\newcommand{\grad}{{\mbox{\boldmath $\nabla$}}}

\begin{document}
\title{Hot Jupiter Magnetospheres}
\author{George B. Trammell, Phil Arras, Zhi-Yun Li}

\altaffiltext{ }{Department of Astronomy, University of Virginia,
P.O. Box 400325, Charlottesville, VA 22904-4325}

%%%%%%%%%%%%%%%%%%%%%%%%%%%%%%%%%%%%%%%%%%%%%%%%%%%%%%%%%%%%%%%%%%%%%

\begin{abstract}

The upper atmospheres of close-in gas giant exoplanets (``hot
Jupiters'') are subjected to intense heating and tidal forces from
their parent stars. The atomic (H) and ionized (H$^+$) hydrogen layers
are sufficiently rarefied that magnetic pressure may dominate gas
pressure for expected planetary magnetic field strength. We examine
the structure of the magnetosphere using a three-dimensional (3D)
isothermal magnetohydrodynamic model that includes: a static ``dead
zone" near the magnetic equator containing gas confined by the magnetic
field; a ``wind zone" outside the magnetic equator in which thermal
pressure gradients and the magneto-centrifugal-tidal effect give rise
to a transonic outflow; and a region near the poles where sufficiently
strong tidal forces may suppress transonic outflow. Using dipole field
geometry, we estimate the size of the dead zone to be several to tens
of planetary radii for a range of parameters.  Tides decrease the size
of the dead zone, while allowing the gas density to increase outward
where the effective gravity is outward. In the wind zone, the rapid
decrease of density beyond the sonic point leads
to smaller densities relative to the neighboring dead zone, which is in
hydrostatic equilibrium.  To understand the appropriate base conditions
for the 3D isothermal model, we compute a simple one-dimensional (1D)
thermal model in which photoelectric heating from the stellar Lyman
continuum is balanced by collisionally-excited Lyman $\alpha$
cooling. This 1D model exhibits a H layer with temperature $T \simeq
5,000-10,000$K down to a pressure $P \sim 10-100$ nbar.  Using the
3D isothermal model, we compute maps of the H column density as well
as the Lyman $\alpha$ transmission spectra for parameters appropriate
to HD 209458b. Line-integrated transit depths $\simeq 5-10\%$ can be
achieved for the above base conditions, in agreement with the results
of Koskinen et al. A deep, warm H layer results
in a higher mass-loss rate relative to that for a more shallow layer,
roughly in proportion to the base pressure. Strong magnetic fields have the
effect of increasing the transit signal while decreasing the mass loss,
due to higher covering fraction and density of the dead zone. Absorption
due to bulk fluid velocity is negligible at linewidths $\ga 100\ {\rm
km\ s^{-1}}$ from line center. In our model, most of the transit signal
arises from magnetically confined gas, some of which may be outside the L1
equipotential. Hence the presence of gas outside the L1 equipotential does
not directly imply mass loss.  We verify {\it a posteriori} that particle mean
free paths and ion-neutral drift are small in the region of interest in
the atmosphere, and that flux freezing is a good approximation. We suggest
that resonant scattering of Lyman $\alpha$ by the magnetosphere may be
observable due to the Doppler shift from the planet's orbital motion,
and may provide a complementary probe of the magnetosphere. Lastly,
we discuss the domain of applicability for the magnetic wind model
described in this paper as well as the Roche-lobe overflow model.

\end{abstract}

%%%%%%%%%%%%%%%%%%%%%%%%%%%%%%%%%%%%%%%%%%%%%%%%%%%%%%%%%%%%%%%%%%%%%

\section{Introduction}

Hot Jupiters, the gas giant exoplanets found at small orbital
radii, $D \la 0.1\ {\rm AU}$, present an opportunity to study planets
orbiting in an extreme environment very close
to their parent stars. They experience insolation levels $\sim 10^4$
times greater than solar system gas giants. As a consequence of the
high temperatures generated by EUV heating, the gas pressure scale
height is comparable to the planetary radius, creating an extended
upper atmosphere of gas potentially observable through transmission
or reflection spectroscopy. Heating may also create an outward pressure
force, complemented by outward centrifugal and
tidal forces, which may drive an outflow leading to mass and angular
momentum loss from the planet.

While most observational effort for the transiting planets has
centered on the lower atmosphere, near the photospheres for optical
and infrared continuum radiation at mbar-bar pressures, there
has also been progress in probing higher levels in the atmosphere.
As these observations probe a larger area surrounding the planet,
the transit depth is correspondingly larger than that from the
photospheric radius, $R_{\rm ph}$. Bound-bound atomic lines in
the transmission spectrum from NaI \citep{2002ApJ...568..377C}, HI
\citep{2003Natur.422..143V, 2004ApJ...604L..69V, 2008A&A...483..933E}, OI
and CII \citep{2004ApJ...604L..69V}, and SiIII \citep{2010arXiv1005.1633L}
have been observed for HD 209458b; HI in HD 189733b
\citep{2010A&A...514A..72L}; and MgII \citep{2010ApJ...714L.222F}, and
possibly other species at lower signal to noise, in WASP-12b. Bound-free
transitions from a population of hydrogen in the $n=2$ state
have been detected in the transmission spectrum of HD 209458b
\citep{2007Natur.445..511B}, although no bound-bound transitions from
$n=2$ were found \citep{2004PASJ...56..655W}.

The large transit depths observed for the Lyman $\alpha$ line in HD
209458b imply neutral H atoms occupy an occulting area equivalent to
an optically thick disk extending out to several planetary radii,
comparable to the Roche-lobe radius for this close-in planet. This
led \citet{2003Natur.422..143V} to suggest that the H atoms are
escaping from the planet, as they would no longer be bound to the
planet outside the Roche-lobe radius.  The presence of heavy atoms (C,
O, etc.), whose scale height should be far smaller than that of the H
atoms, was suggested to be due to a combination of efficient turbulent
mixing and drag forces from the escaping gas, entraining the heavier
elements in an outflow \citep{2004ApJ...604L..69V, 2007P&SS...55.1426G}.
\citet{2007ApJ...671L..61B} used archival HST data to rederive the transit
depth, and challenged the claim that atmospheric escape was occurring,
due to a derived transit depth that would imply the H atoms are within the
Roche lobe radius.  Two subsequent studies \citep{2008ApJ...688.1352B,
2008ApJ...676L..57V} elucidated the dependence of transit depth on
the analysis method and wavelength range studied. The recent study by
\citet{2010arXiv1005.1633L} attempts to constrain the wind velocity in
HD 209458b using the line profiles for the CII and SiIII lines.

Most theoretical efforts to explain the aforementioned observations
invoke EUV heating to temperatures $\sim 10^4\ {\rm K}$, creating a
thermally driven, possibly transonic hydrodynamic outflow from the planet
\citep{ 2004Icar..170..167Y, 2006Icar..183..508Y, 2005ApJ...621.1049T,
2007P&SS...55.1426G,2009ApJ...693...23M, 2009ApJ...694..205S}. While
initiated by outward gas pressure gradients, a thermally-driven wind
can be significantly accelerated by the tidal gravity\footnote{In this
paper we will use the phrase``tidal gravity" to include the effect
of both the stellar tide and the centrifugal force, assuming synchronized
rotation (see \S~\ref{sec:U}.)}. An alternative model is to 
treat atmospheric escape as Roche-lobe overflow \citep{2003ApJ...588..509G,
2010Natur.463.1054L,2010arXiv1005.4497L}, assuming that the fluid
can only reach the sound speed in the vicinity of the L1 Lagrange
point.
% Presumably, the appropriate model can be chosen by a comparison
% of the Roche lobe radius and the sonic point radius.
%\citep{2008Natur.451..970H,
%2010ApJ...709..670E, 2010Natur.463.1054L,2010SoSyR..44...96S}
%\citep{2009ApJ...693..868K,2010arXiv1004.1396K}

%For gas giants, H is the most abundant element and heating is
%dominated by photoionization of atomic H in the thermosphere
%\citep{2004Icar..170..167Y}. Photoionization leads to significantly
%ionized gas at high altitudes that is subject to magnetic forces. In
%light of the high ionization fraction, the planetary magnetic field, due
%to dynamo action in the planet's core, may be dynamically important and
%affect density and velocity profiles in the upper atmosphere.  We will
%show that for reasonable field strengths, even far smaller than Jupiter's
%(equatorial field $B_{\rm J,eq}=4.3\ {\rm G}$), the magnetic pressure
%dominates gas pressure in the bulk of the atomic and ionized layers. As
%a result, the gas is not free to escape at all latitudes, as discussed
%by \citet{2004Icar..170..167Y}.  Rather, the gas can only escape along
%field lines too weak to confine it.
For gas giants, H is the most abundant element and heating is
dominated by photoionization of atomic H in the thermosphere
\citep{2004Icar..170..167Y}.  Photoionization leads to significantly
ionized gas at high altitudes that is subject to magnetic forces. In light
of the high ionization fraction, the planetary magnetic field, due to
dynamo action in the planet's core, may be dynamically important and
affect density and velocity profiles in the upper atmosphere. This paper
is a first attempt to include the effects of the planet's magnetic
field combined with tidal and centrifugal forces on upper atmosphere
structure for hot Jupiters.

%Within this context of a magnetohydrodynamic (MHD) outflow,
%the result of a magnetized, hot, rotating surface is the 
%``helmet streamer" configuration for the magnetic field that is
%a classic feature of MHD winds from an isolated object (see Figure
%\ref{fig:cartoon}).  The polar region supports an outflow while the
%equatorial region contains static, magnetically confined gas. But
%for planets sufficiently close to their parent star, the L1 Lagrange
%point is inside this dead zone, and the gas density can actually increase
%outward, supported by the magnetic field, since the effective gravity
%is outward. This paper is a first attempt to include the effects of the
%planet's magnetic field combined with tidal and centrifugal forces on
%upper atmosphere structure for hot Jupiters.

\begin{figure}[t]
\epsscale{1.2}
\plotone{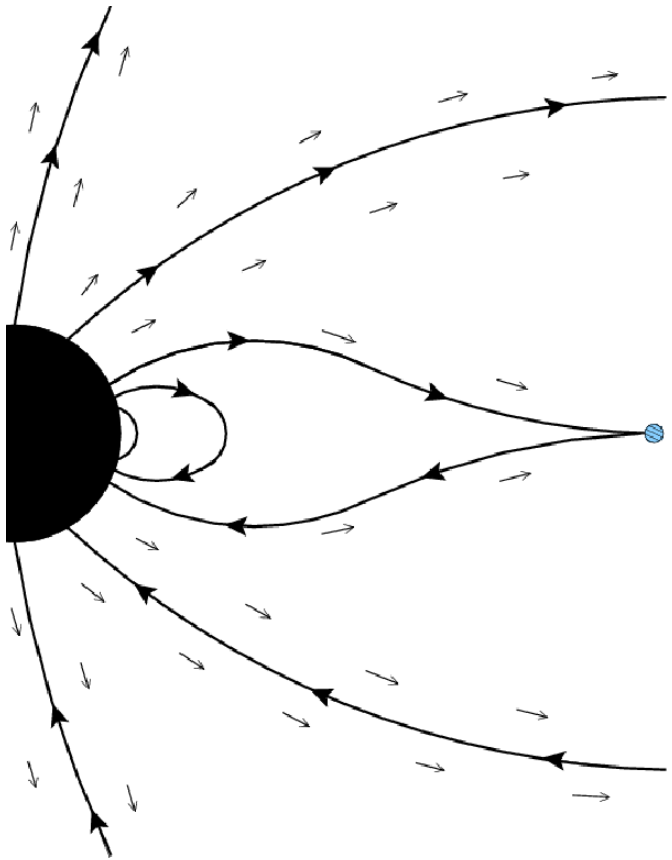}
\caption{
Schematic model for magnetic field structure (thick lines and arrows)
and fluid velocity (thin arrows).  In the polar ``wind zone", magnetic
field lines are open allowing outflow.  In the equatorial ``dead zone"
close to the planet, the gas has zero velocity and no outflow occurs. The
dead zone ends in a cusp-type neutral point, denoted by the dashed circle
at several planetary radii, outside of which field lines are open at
all angles. This figure is characteristic of the weak tide limit, while
in the strong tide limit the polar wind would be partially suppressed
(see \S~\ref{sec:wind}).
}
\label{fig:cartoon}
\end{figure}

We model the magnetized, hot, rotating portion of the planet's
upper atmosphere in the context of magnetohydrodynamic (MHD)
outflows, the theory of which was originally developed for stars
and accretion disks. Figure \ref{fig:cartoon} shows a cartoon of
the expected structure for the wind from an isolated object with a
surface dipole field: the polar region supports an outflow while the
equatorial region contains static, magnetically confined gas \citep[see
also][]{2004Icar..170..167Y}. This configuration applies interior to
the magnetosphere-stellar wind interaction, for planets not too near
the parent star. Near the parent star, tidal forces are an important
consideration, and we will show that tides may strongly affect the
density profile and size of the static region, and may even shut the
wind off in the polar region.

In the first half of the paper (\S~\ref{sec:helmet} through
\S~\ref{sec:wind}), we develop a general theory of isothermal hot Jupiter
magnetospheres. Section \ref{sec:helmet} discusses the problem setup
and approximations used to obtain a solution. Planetary magnetic
field strengths predicted by dynamo models are discussed in section
\ref{sec:B}. Section \ref{sec:pi} discusses qualitative features of
atmospheric structure, and motivation for the existence of a dead zone.
Section \ref{sec:U} reviews the centrifugal and tidal forces, and
discusses the projection of these forces along magnetic field lines. The
structure of the dead zone is discussed in section \ref{sec:dead},
followed by the wind zone in section \ref{sec:wind}.

%The second half of the paper (\S~\ref{sec:Hlayer} through
%\S~\ref{sec:lymanalpha}) is devoted to confronting the theory with
%transit observations. In section \ref{sec:Hlayer}, we construct a
%1D thermal model to yield appropriate values for the base pressure
%and temperature for the 3D global models that we present in section
%\ref{sec:global}. Mass loss rates and spin-down torques are computed in
%section \ref{sec:mdot}.  Neutral H column density maps for the global
%models are used to illustrate the dependence on key parameters in section
%\ref{sec:NH}. Lyman $\alpha$ transmission spectra in comparison with
%observations and scattering of stellar Lyman $\alpha$ by the planet are
%presented in section \ref{sec:lymanalpha}. Finally, we compare and
%contrast our magnetic wind model with the standard Roche-lobe overflow
%model in section \ref{sec:roche}.  Summary and discussion are given
%in section \ref{sec:summary}.  The MHD wind equations and ion-neutral
%coupling are discussed in Appendices \ref{sec:mhd} and \ref{sec:mfp},
%respectively.
The second half of the paper (\S~\ref{sec:Hlayer} through
\S~\ref{sec:lymanalpha}) uses the Lyman $\alpha$ transmission spectra to
constrain parameters of the global 3D models. In section \ref{sec:Hlayer},
we construct a 1D model in hydrostatic, thermal, and ionization balance
to compute appropriate values for the base pressure and temperature for
the 3D global models that we present in section \ref{sec:global}. Mass
loss rates are computed and spin-down torques are estimated, in section
\ref{sec:mdot}.  Neutral H column density maps for the global models
are used to illustrate the dependence on key parameters in section
\ref{sec:NH}. Lyman $\alpha$ transmission spectra in comparison with
observations and scattering of stellar Lyman $\alpha$ by the planet
are presented in section \ref{sec:lymanalpha}. Finally, we compare and
contrast our magnetic wind model with the standard Roche-lobe overflow
model in section \ref{sec:roche}.  Summary and discussion are given
in section \ref{sec:summary}.  The MHD wind equations and ion-neutral
coupling are discussed in Appendices \ref{sec:mhd} and \ref{sec:mfp},
respectively.

%%%%%%%%%%%%%%%%%%%%%%%%%%%%%%%%%%%%%%%%%%%%%%%%%%%%%%%%%%%%%%%%%%%%%

\section{ problem setup and approximations }
\label{sec:helmet}

In this section we outline the problem to be solved, and the simplifying
assumptions used to find solutions. Except for section \ref{sec:Hlayer},
this paper discusses a 3D isothermal model of the upper atmosphere. The
isothermal model is parametrized by an effective sound speed, as well
as the pressure at a fiducial radius.  Appropriate values for these
parameters are discussed in a simple 1D spherical model in section
\ref{sec:Hlayer}, including photoionization and thermal equilibrium.

In the 3D model, we compute approximate solutions of the one-fluid
MHD equations with an inner boundary at the base of the warm H layer,
and an outer boundary which extends to at least ten planetary radii
(or one stellar radius). We treat the gas as having constant isothermal
sound speed $a=\sqrt{k_bT/\mu m_p}$, where $T$ is gas temperature of the
fluid, $\mu$ is the mean molecular weight, $k_b$ is Boltzmann's constant,
and $m_p$ is the proton mass. At the inner boundary, the density and
pressure are assumed to follow equipotentials.  We assume the planet's
rotation rate is synchronized with its orbital motion around the parent
star, giving orbital and spin angular velocity $\Omega$. We work in a
coordinate system centered on the planet and rotating at rate $\Omega$.
The stellar gravity is included in the tidal approximation, and an
effective potential $U$ is composed of the planetary gravity, stellar
gravity, and the centrifugal force. We specify a specific magnetic field
geometry which is dipole near the planet and matches onto a radial field
at the dead zone radius.  We compute the ionization fraction with a
simple, optically thin model applied to the derived gas densities of the
MHD model.  To summarize, we have made several simplifying assumptions
in order to focus on the new physics arising from MHD effects.  We now
discuss these simplifying approximations in more detail.

The simultaneous inclusion of photoionization heating, chemical reactions
and collisional coupling between different species, stellar tidal forces,
and the simultaneous interaction with the stellar wind in the presence of
magnetic field is a formidable problem. Our approach is to first ignore
the interaction with the stellar wind, but to include the effect of the
stellar tidal forces felt by the planet's atmosphere. The interaction with
the stellar wind may alter the results of this paper in several ways 
%(see e.g.{\red ,}\citealt{2009ApJ...693...23M,2009ApJ...694..205S}). 
\citep[see e.g.,][]{2009ApJ...693...23M,2009ApJ...694..205S}.
The stellar wind
will limit the size of the magnetosphere, as determined by stress balance
at the magnetopause %(e.g. \citealt{2007P&SS...55..589P}).
\citep[e.g.,][]{2007P&SS...55..589P}.
Reconnection between field lines in the stellar wind
and magnetosphere may lead to magnetospheric convection, limiting the 
high density region to be inside a plasmapause \citep{2004pspi.book.....P}, 
as for Earth.
Finally, reconnection may also generate non-thermal plasma populations.
We do not consider the interaction with the stellar wind in order to construct
the simplest possible model. 
%This approach may
%fail for the open field lines subject to ``magnetospheric convection"
%motions due to reconnection between the planet and stellar wind field
%(see, e.g. \citealt{1995isp..book.....K}).

Another key approximation is that we treat photoionization heating as
being spherically symmetric, creating a hot layer uniformly over the
planet. In reality, the night side temperature and ionization state may
depend on day-night heat redistribution, and downward heat conduction
along field lines not in the planet's shadow.  In perfect MHD, such
redistribution would be highly constrained in the magnetically dominated
upper atmosphere, but finite conductivity may allow field lines to slip
through the gas \citep{1959JGR....64.1219G}.  Even on the day side,
large gas density outside the Roche lobe may project a non-spherically
symmetric shadow on the deeper layers.

%Near the planet, the dynamo-generated, roughly dipole field from the
%planet's core is expected to dominate. The field lines become combed
%into a nearly radial direction beyond the dead zone radius. Such
%a geometry has been used before in the context of the stellar wind
%\citep{1968MNRAS.138..359M,1987MNRAS.226...57M,1974MNRAS.166..683O} and
%we will adopt it here. This field geometry will be implemented
%in the global models presented in section 9, which is motivated by the
%discussion in Appendix A.
Near the planet, the dynamo-generated, roughly dipole field from
the planet’s core is expected to dominate. Moving outward,
currents generated in the magnetosphere comb the field lines
into a nearly radial direction beyond the dead zone radius. Such
a geometry has been used before in the context of the stellar wind
\citep{1968MNRAS.138..359M,1987MNRAS.226...57M,1974MNRAS.166..683O} and
we will adopt it here. This field geometry will be implemented in the
global models presented in section 9, and is motivated in the discussion
of Appendix A.

Finally, the one-fluid approximation assumes that mean free paths are
sufficiently small that relative motion of different species can be
ignored.  In Appendix \ref{sec:mfp} we will check this assumption {\it
a posteriori} for our models, which estimate particle densities and
velocities in the dead and wind zones. Specifically, we will show that
the electron-proton-hydrogen atom gas is well coupled collisionally for
the parameters of interest, and therefore the drift velocity is small and
hydrogen atoms have short mean free paths in the dead zone region. As a
consequence, neutral hydrogen atoms do not fly ballistically through the
magnetosphere, and photoionization equilibrium is a good approximation
when computing the ionization fraction. Further, we compute the rate at
which magnetic field can drift relative to the fluid. For this thermal
population of particles in the magnetosphere, we find that the dead zone
gives the largest observable transit signal, and that the bulk of the
hydrogen atoms in the dead zone are not escaping.

In the next section we will review expected magnetic field strengths
for hot Jupiters.

%%%%%%%%%%%%%%%%%%%%%%%%%%%%%%%%%%%%%%%%%%%%%%%%%%%%%%%%%%%%%%%%%%%%%

\section{ expected magnetic field strengths }
\label{sec:B}

The importance of the magnetic field for the upper atmosphere depends
critically on the field strength. However, the magnetic fields of hot
Jupiters are currently unconstrained by observation. This section will
use theoretical considerations to estimate likely field strengths for
 hot Jupiters.

\citet{2004ApJ...609L..87S} computed that Rayleigh numbers in hot
Jupiters are typically much larger than the critical Rayleigh number for
thermal convection in the metallic core.  Using estimates of the fluid
velocity carrying the heat flux, he found that the magnetic Reynolds
number is much larger than unity, and that dynamo action can occur.
%Further, the rotation of synchronized planets is sufficiently rapid
%for the Rossby number $Ro$ to be $\la 10^{-3}$.  
He argued that if the dynamo operated with Elsasser number of order
unity, then $B \sim (2 \rho \Omega \lambda_B)^{1/2} $, where $\rho$
is the mass density and $\lambda_B$ is the magnetic diffusivity. The
dominant scaling important for hot Jupiters is then with rotation: $B
\propto \Omega^{1/2}$. For synchronized planets with orbital periods of
a few days, this scaling predicts that the field for hot Jupiters should
be smaller than that of Jupiter (equatorial field $B_{\rm J,eq}=4.3\
{\rm G}$) by a factor of a few.

The opposite conclusion may be drawn from the recent results of
\citet{2009Natur.457..167C}.  %, as we now explain.  They argue that
As the rotation rate is increased above a critical value, the field
strength no longer increases with rotation rate, and dynamo simulations
give a magnetic field strength $B \sim (\rho F_{\rm core}^2)^{1/3}$,
nearly independent of rotation rate and magnetic diffusivity, where
$F_{\rm core}$ is the heat flux  escaping from the conducting core.
\citet{2009Natur.457..167C} show that this scaling applies to both planets
and rapidly rotating, low mass stars over many orders of magnitude in heat
flux. They argue that the dependence on $F_{\rm core}$ arises since it
is the heat flux reservoir that sustains the magnetic
field against Ohmic dissipation.  To be in the saturated regime, the
Rossby number $Ro$ must satisfy $Ro=V_{\rm ed}/\Omega \ell \la 0.1$, where
$V_{\rm ed}$ is the typical velocity of the eddies transporting heat, and
$\ell$ is the size of the conducting region; $\ell \sim R$ for $M_p \ga
0.5M_J$. \citet{2004ApJ...609L..87S} estimated synchronized planets in few
day orbits to have $Ro \ll 0.1$, using heat fluxes comparable to that of
Jupiter. Hence these planets are expected to be in the saturated regime.

The large radii of hot Jupiters are currently not well understood, since
the cooling and contraction time for passively cooling planets, even
allowing for irradiation by the star, is far shorter than the age for a
number of observed planets \citep[e.g.,][]{2009SSRv..tmp..115F}. This
has led to the suggestion that these planets are not passively cooling,
but rather have an anomalous source of internal heating, which is as
yet unidentified but balances the core cooling rate.  To assess the
required heating rates, \citet{2009arXiv0901.0735A} computed cooling
flux from the core for planets as a function of radius 
%(their Figure
%11; see \citealt{2006ApJ...650..394A} for a discussion of the cooling
%luminosity of irradiated hot Jupiters).  
\citep[their Figure 11; see][for a discussion of the cooling luminosity
of irradiated hot Jupiters]{ 2006ApJ...650..394A}.  For Jupiter-mass
planets in the radius range $R_{\rm ph}=1.3-1.5 R_J$, the cooling flux
is larger than that of Jupiter by a factor $10^2-10^3$, for which the
\citet{2009Natur.457..167C} scaling would give magnetic fields $5-10$
times larger than Jupiter. Larger mass planets with the same radius would
have larger cooling fluxes, and vice versa. For instance, WASP 12b,
WASP 17b and TRES 4 have radii $R_{\rm ph} \sim 1.8R_J$ and masses in
the range $0.5-1.5M_J$\footnote{http://exoplanet.eu/catalog-transit.php}, for
which the cooling fluxes would be $10^3-10^4$ times higher than Jupiter,
implying fields larger than Jupiter by factors of $10-20$.

In summary, the recent results on dynamo theory from
\citet{2009Natur.457..167C},  and the assumption that hot Jupiter cores
are subject to an externally powered heating \citep{2009arXiv0901.0735A},
argue that field strengths may be up to an order of magnitude larger
than that of Jupiter. We will take this as motivation to explore a wide
range of parameter space for the magnetic field in our calculations.

In the next section, we show that the photoionized H and H$^+$ layers
are magnetically dominated for field strengths comparable to Jupiter or
Saturn, and motivate the existence of a dead zone by a toy problem.

%%%%%%%%%%%%%%%%%%%%%%%%%%%%%%%%%%%%%%%%%%%%%%%%%%%%%%%%%%%%%%%%%%%%%

\section{dead zone-wind zone structure of the upper atmosphere }
\label{sec:pi}

\citet{2004Icar..170..167Y} and \cite{2007P&SS...55.1426G} presented
detailed calculations of the transition between the molecular lower
atmosphere (H$_2$), the layer dominated by atomic hydrogen (H), and the
ionized upper atmosphere (H$^+$). In this paper, we restrict attention to
the H and H$^+$ regions, where the transmission spectrum is formed. The
strong heating in these layers due to UV photon energy deposition
raises the temperature to $T
\simeq 10^4\ {\rm K}$. As a consequence of the increased temperature and
low mean molecular weight, the radial extent of the H and H$^+$ layers
($\ga R_J$) is expected to be much larger than that of the H$_2$ layer
above the photosphere ($\la (0.1-0.2)\times R_J$).

We define the lower boundary of our wind model to be at the base of
the warm ($T \ga 5000\ {\rm K}$) H layer, at base radius $R$ and base
pressure $P_{\rm base}$. The base of the warm layer is a crucial parameter
for the transit depth.  As discussed in the phenomenological model of
\citet{2010arXiv1004.1396K}, the transit depth of HD 209458b could be
understood as being due to thermal $\simeq 10^4\ {\rm K}$ H gas extending
down to $\sim 10-100\ \rm nbar$ pressures.  In section \ref{sec:Hlayer},
we compute a simple 1D model for ionization and thermal equilibria in
the H and H$^+$ layers which shows that such base conditions are indeed
possible.

The Lyman $\alpha$ transmission spectrum of HD 209458b shows absorption
by the planetary atmosphere at linewidths $\Delta v \ga 100\ \rm
km\ s^{-1}$ from line center. At this linewidth, the cross section
is $\sigma_\nu \simeq 2 \times 10^{-19}\ \rm cm^2$ (see Figure
\ref{fig:lymanalphaxsec}). Optical depth unity requires a hydrogen column
$N_H \simeq 1/\sigma_\nu \simeq 5 \times 10^{18}\ \rm cm^{-2}$. Assuming
the gas is dominated by atomic hydrogen, the pressure at this level in
the atmosphere is $P \simeq gm_p N_H \simeq 2\ {\rm nbar}\ (g/300\ \rm cm\
s^{-2})$ (see Figure \ref{fig:pieq}).  This is a factor $\sim 10^8$ more
rarefied than the optical photosphere for continuum radiation at pressure
$P_{\rm ph} \simeq {\rm 100\ mbar}$. Magnetic forces dominate in this
layer if $B^2/8\pi \ga 2\ {\rm nbar}\ (g/300\ \rm cm\ s^{-2}) $, implying
a critical field strength $B_{\rm crit}\ga 0.25\ {\rm G}\ (g/300\ {\rm
cm^2\ s^{-1}})^{1/2}$. This is less than Jupiter's equatorial magnetic
field $B_{\rm J,eq}=4.3\ {\rm G}$ and comparable to Saturn's equatorial
field $B_{\rm S, eq}=0.22\ {\rm G}$. Moving upward, if the gas pressure
drops much faster than magnetic pressure, the atmosphere can become
highly magnetically dominated --- a magnetosphere.  
%and magnetic effects may dominate the structure of the atmosphere.  
%See Figure \ref{fig:pvsr} for a comparison of gas and magnetic pressure versus radius.

The theory of thermally and magneto-centrifugally driven MHD winds gives
guidance on the upper atmosphere structure in the magnetically dominated
case (for a good review see \citealt{1996astro.ph..2022S}). Consider
a thought experiment in which a non-magnetic spherically symmetric wind with velocity $v_\infty$ and mass loss rate
$\dot{M}$ exists at time $t<0$, and at time $t=0$ a dipole magnetic field
is turned on. On which field lines can the wind overpower the magnetic
forces and open the field lines to infinity?
%The field lines with footpoint at colatitude $\theta_b$
%extend along $r(\theta)=R\sin^2\theta/\sin^2\theta_b$. 
%Hence 
The magnetic pressure on the equator ($\theta=\pi/2$) is
weaker than at the footpoint (at angle $\theta_b$) by a factor
$[B(\pi/2)/B(\theta_b)]^2=[R/r(\pi/2)]^6 =\sin^{12}\theta_b$ (see
eq.\ref{eq:fieldline}).  This powerful dependence on footpoint position
means there are always field lines near the magnetic poles which open to
infinity on which a wind can outflow (see Figure \ref{fig:cartoon} for a
cartoon). The reason is that at the equator, the wind ram pressure can
overcome the steeply falling magnetic pressure at sufficiently large
distance from the planet.  If we assume the wind
ram pressure decreases outward as $\rho v^2=\dot{M}v_\infty/4\pi r^2$,
then the critical footpoint angle inside of which a ``polar wind" occurs
is $\sin\theta_b \simeq \left( \dot{M}v_\infty/R^2B_0^2 \right)^{1/8}$.
For fiducial polar magnetic field $B_0=8.6\ {\rm G}$, $R=1.4R_J$,
(constant) flow speed $v_\infty=10\ {\rm km\ s^{-1}}$ and mass loss
rate $\dot{M}=10^{11}\ {\rm g\ s^{-1}}$ (motivated by the studies of
\citealt{2004Icar..170..167Y, 2007P&SS...55.1426G, 2009ApJ...693...23M}),
we find the polar cap size occupied by open field lines is $\theta_b
\simeq 14^\circ$, and the last closed field lines is at equatorial radius
$r(\pi/2) \simeq 16 R$. This static, closed field line region is referred
to as the ``dead zone" \citep{1968MNRAS.138..359M}.  These estimates
suggest that the region within a few planetary radii, where the transit
signal arises, is filled primarily with static gas in the dead zone,
rather than outflowing gas, as has been previously assumed.

For close-in, tidally-locked planets, tides and centrifugal forces play
a key role in upper atmosphere structure.  In the next section we review
the effective potential near the planet, and the projection of forces
along dipole field lines.

%%%%%%%%%%%%%%%%%%%%%%%%%%%%%%%%%%%%%%%%%%%%%%%%%%%%%%%%%%%%%%%%%%%%%

\section{tidal force and magnetic geometry }
\label{sec:U}

Gas in the upper atmosphere of the planet is subject to three
potential forces: gravity from the planet, tidal gravity from the
star, and the centrifugal force due to the planetary rotation. For a
synchronized planet, the centrifugal and tidal forces, or just tidal
force for short, are comparable in strength, although their angular
dependence is different. For position vector
$\vec{x}=(r,\theta,\phi)$ relative to the center of the planet, and star
at position $\vec{x}_\star=(D,\pi/2,0) = D\vec{e}_x $, the sum of the
three accelerations is
\be
\vec{a}(\vec{x}) & =& - \grad U(\vec{x}),
\ee
where the effective potential is given by
\be 
U(\vec{x}) & = & - \frac{GM_p}{|\vec{x}|} - \frac{GM_\star}{|\vec{x}-\vec{x}_\star|}
+ \frac{GM_\star \vec{x} \cdot \vec{x}_\star}{|\vec{x}_\star|^3}
- \frac{1}{2} \left| \vec{\Omega}\times \vec{x} \right|^2
\label{eq:U_0} \\ 
& \simeq & -  \frac{GM_p}{r} - \frac{1}{2} \Omega^2 r^2
\left( f \sin^2\theta - 1 \right).
\label{eq:U}
\ee
Equipotentials are shown in Figure 6.2 of \citet{1978ASSL...68.....K}. The
longitude-dependent function $f=1 + 3 \cos^2\phi$. The vector
angular velocity of the orbit is $\vec{\Omega}=\Omega \vec{e}_z$,
where $\vec{e}_z$ is normal to the orbital plane, and $\Omega
= [G(M_\star+M_p)/D^3]^{1/2}$.  The first and second terms
in eq.\ref{eq:U_0} are the potential of the planet and star,
respectively. The third term in eq.\ref{eq:U_0} is due to the motion of
the origin of the coordinate system. The last term in eq.\ref{eq:U_0}
is due to the centrifugal force. It may be shown that eq.\ref{eq:U_0}
is equivalent to the usual Roche potential with origin at the center of
mass by combining the third and fourth terms.  The form in eq.\ref{eq:U}
is an expansion in the limit $r \ll D$, and agrees with that for ``Hill's
limit" found in \citet{2000ssd..book.....M} when evaluated in the orbital
plane ($\theta=\pi/2$).

The accelerations are given by
\be 
a_r & = & - \frac{\partial U}{\partial r}
=  
- \frac{GM_p}{r^2} + \Omega^2 r \left( f\sin^2\theta - 1 \right), \label{eq:ar}
\\
a_\theta & =& - \frac{1}{r} \frac{\partial U}{\partial \theta}
= f \Omega^2 r \sin\theta \cos\theta
\\
a_\phi & = & - \frac{1}{r \sin\theta} \frac{\partial U}{\partial \phi}
= - 3 \Omega^2 r \sin\theta \sin\phi \cos\phi.
\ee
The vector acceleration $\vec{a}$ should not be confused with the sound
speed $a$. If we denote the coordinate along the star-planet line $x =
r\sin\theta\cos\phi$, and normal to the orbit plane $z=r\cos\theta$, then
the tidal force is outward when $3x^2>z^2$, i.e. within latitudes $-\pi/3$
to $\pi/3$ of the equator. The tidal force is zero in the $y$-direction.
Hence when observing a planet in the plane of the sky during transit,
the tidal forces are inward along $\vec{e}_z$, zero along $\vec{e}_y$, and
away from the planet along $\vec{e}_x$, the line of sight during transit.

%The $a_\theta$ acceleration is always toward the equator and is
%stronger along the star-planet line. The effects of tides and centrifugal
%force add together in $a_\theta$, and have the same angular and radial
%dependence, differing only in their longitude dependence. In the absence
%of the tidal force, the centrifugal force acts to funnel gas toward the
%equator, giving rise to a fast polar wind and slow equatorial wind.
%In the dead zone, $a_\theta$ is crucial in determining the radius where the
%force parallel to the loop points away from the planet, giving rise to
%enhanced density at looptops. 

%\begin{figure}[t]
%\epsscale{1.2}
%\plotone{rroche_over_rbase_vs_porb.ps}
%\caption{ 
%Ratio of distance to the L1-L2 Lagrangian points to measure transit
%radius. The open circles, X's and open triangles are planets with
%$M_p/M_J>2$, $1/2 \leq M_p/M_J \leq 2$ and $M_p/M_J<1/2$, respectively.
%Values of $D$, $M_\star$, $M_p$ and $R_{\rm ph}$ are taken from the
%database of transiting exoplanets (http://exoplanet.eu/).
%}
%\label{fig:roche}
%\end{figure}

The L1 and L2 Lagrangian points at radii $r_{\rm L1}$ and $r_{\rm L2}$ are found
by using eq.\ref{eq:ar} along the star-planet line ($\theta=\pi/2, \phi=0$),
giving
\be
r_{\rm L1} \simeq r_{\rm L2} \simeq r_{\rm L}  & \equiv &   D \left(M_p/3M_\star \right)^{1/3} = (GM_p/3\Omega^2)^{1/3}.
\label{eq:rL}
\ee
%Note that for exoplanets, $M_{\rm p}/M_{\star} \ll 1$ in
%eq.\ref{eq:rL} and $r_{\rm L1} \simeq r_{\rm L2}$. The photospheres
%of the observed planets are inside the L1-L2 radii.
The near equality $r_{\rm L1} \simeq r_{\rm L2}$ is due to $M_{\rm p}/M_{\star} \ll 1$
in eq.\ref{eq:rL}. The photospheres of the observed planets are inside the L1-L2 radii.

How does the magnetic field alter the radius beyond which the net gravity
points outward? What is needed is the projection $a_\parallel=\vec{a}
\cdot \vec{b}$ along field lines, where $\vec{b}=\vec{B}/B$ is the
unit vector along the magnetic field direction. In dipole geometry,
approximately correct near the planet,
\be
\vec{B} & = & B_0 \left( \frac{R}{r} \right)^3 \left( \vec{e}_r \cos\theta +
\vec{e}_\theta \frac{\sin\theta}{2} \right).
\label{eq:dipole}
\ee
and the unit vector is
\be
\vec{b} & = & \frac{1}{N} \left( \vec{e}_r \cos\theta  +
\vec{e}_\theta \frac{\sin\theta}{2} \right)
\label{eq:b}
\ee
where the normalization factor is 
\be
N & = & \sqrt{\cos^2\theta + \sin^2\theta/4}
= \sqrt{1-3\sin^2\theta/4}.
\ee
The parallel acceleration is then
\be
a_\parallel & =& - \vec{b} \cdot \grad U
= \frac{\cos\theta}{N} \left[ - \left( \frac{GM_p}{r^2} + \Omega^2 r \right)
+ \frac{3}{2} f \Omega^2 r \sin^2 \theta \right].
\label{eq:apar}
\ee

The quantity in brackets in eq.\ref{eq:apar} must be positive in order
for the net acceleration to be away from the planet. To solve for the
radius at which the net acceleration is zero, we express $\theta$ in
terms of $r$ along dipole field lines using
\be
r(\theta) & = & R \frac{\sin^2\theta}{\sin^2\theta_b} \equiv r_{\rm eq} \sin^2\theta
\label{eq:fieldline}
\ee
where
\be
r_{\rm eq} & = & R/\sin^2\theta_b
\label{eq:req}
\ee
is the equatorial radius of the field line with footpoint at
$\theta_b$. The ``magnetic Roche lobe radius", $r_{\rm RB}$, at which
the projected acceleration $a_\parallel=0$ is given by the solution of
the equation
\be
\frac{GM_p}{r_{\rm RB}^2} + \Omega^2 r_{\rm RB}  & = &
\frac{3}{2} f \Omega^2 \frac{r_{\rm RB}^2}{r_{\rm eq}}.
\label{eq:rRB}
\ee
Solutions can only exist in the range of radii $2r_{\rm eq}/3f \leq
r_{\rm RB} \leq r_{\rm eq}$.  Solutions for $r_{\rm RB}$ exist first at
the looptop $r_{\rm RB}=r_{\rm eq}$ for loops of critical size
\be
r_{\rm eq, crit} & = & \frac{r_{\rm L}}{(f/2-1/3)^{1/3}}.
\label{eq:reqcrit}
\ee
That is, when the loop size becomes larger than about the L1-L2
distance, the outer part of the loop can have net acceleration
pointing away from the planet. Note that this statement applies
even in the plane where $\cos^2\phi=0$, where the radial component
of the tidal force points inward. The magnetic geometry allows a
``magnetic Roche lobe" $r_{\rm RB} \sim r_{\rm R}$ to exist at all
longitudes, in contrast to the unmagnetized case. 
%This fact will
%be important for both the density and wind velocity profiles.

In the next section we model the gas density in the dead zone.

%%%%%%%%%%%%%%%%%%%%%%%%%%%%%%%%%%%%%%%%%%%%%%%%%%%%%%%%%%%%%%%%%%%%%

\section{ the dead zone }
\label{sec:dead}

The 3D MHD wind equations are presented in Appendix \ref{sec:mhd}. There
we derive the Bernoulli constant along field lines and discuss how gas
pressure discontinuities at the dead zone - wind zone boundaries give
rise to current sheets which alter the magnetic field configuration. For
the present section which discusses the dead zone, the main concept needed is
that hydrostatic balance applies along field lines.  We will approximate
the field lines as dipolar.

In the dead zone, the velocity along field lines $\vec{v}=0$. Setting
$\vec{v}=0$ in eq.\ref{eq:mom} and dotting this equation with $\vec{b}$
to eliminate the Lorentz force we find the equation of hydrostatic
balance along field lines
\be
\frac{1}{\rho} \frac{dP}{ds} & = & a^2 \frac{d\ln\rho}{ds} = - \frac{dU}{ds}
\label{eq:hb}
\ee 
where $d/ds=\vec{b} \cdot \grad$ is the derivative along field lines.
Under the isothermal assumption, eq.\ref{eq:hb} can then be integrated
to give the run of pressure and density along a field line with base
position $(\theta_b,\phi)$:
\be
\frac{P(r,\theta,\phi)}{P(R,\theta_b,\phi)} & = & 
\frac{\rho(r,\theta,\phi)}{\rho(R,\theta_b,\phi)} 
\label{eq:isoP0}
\\ & =  &
\exp\left[ - \left( \frac{U(r,\theta,\phi)-U(R,\theta_b,\phi)}{a^2} \right) \right]. \nonumber
%\label{eq:isoP0}
\ee
We treat the density and pressure at the base as being
along equipotentials. Defining $\rho_{ss}=P_{ss}/a^2 =
\rho(r=R,\theta=\pi/2,\phi=0)$ to be the value at the substellar point
at the base radius, the density at the base radius at other points is
\be
\rho_b(\theta_b,\phi) & = & \rho_{ss} \exp \left[ - \left( \frac{U(R,\theta_b,\phi)-U(R,\pi/2,0)}{a^2}
\right) \right].
\label{eq:rhob}
\ee
Combining eq.\ref{eq:isoP0} and \ref{eq:rhob} then gives
\be
\frac{P(r,\theta,\phi)}{P_{ss}} & = &
\frac{\rho(r,\theta,\phi)}{\rho_{ss}}
\label{eq:isoP}
\\ & =  &
\exp\left[ - \left( \frac{U(r,\theta,\phi)-U(R,\pi/2,0)}{a^2} \right) \right]. \nonumber
%\label{eq:isoP}
\ee
Note that because we have assumed the density and pressure surfaces at
the base are along equipotentials, eq.\ref{eq:isoP} satisfies
\be
0 & = & - \grad P - \rho \grad U
\label{eq:hb3d}
\ee
in all three directions, not just along field lines. In this case $-
\grad U$ is balanced by gas pressure forces due to a non-spherical
distribution of mass. If, on the other hand, temperature, density or
pressure surfaces were not along equipotentials, then eq.\ref{eq:hb3d}
would not be satisfied perpendicular to field lines, and the trans-field
force balance (eq.\ref{eq:Jperp}) would be required to understand the
required currents.

\begin{figure}[t]
\epsscale{1.2}
\plotone{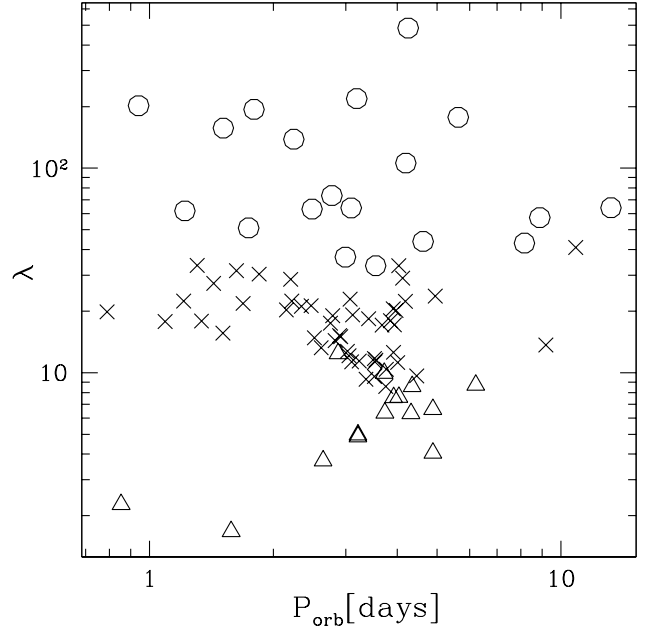}
\caption{
Values of $\lambda$ calculated using eq.\ref{eq:lambda} for the current
database of transiting exoplanets.  The parameters used are $a=9.3\
{\rm km\ s^{-1}}$, appropriate to $T=10^4\ {\rm K}$, $\mu=1$, and
$R=R_{\rm ph}$.  The open circles, X's and open triangles are planets with
$M_p/M_J>2$, $1/2 \leq M_p/M_J \leq 2$ and $M_p/M_J<1/2$, respectively.
Values of $P_{\rm orb}$, $M_p$ and $R_{\rm ph}$ are taken from the
current database of transiting exoplanets (http://exoplanet.eu/).
}
\label{fig:lambdaobs}
\end{figure}

The potential difference in eq.\ref{eq:isoP} can be written in 
dimensionless form
\be
&& \frac{U(r,\theta,\phi)-U(R,\pi/2,0)}{a^2}
\\ & = & 
\lambda \left( 1 - \frac{R}{r} \right)
+ \frac{1}{2} \epsilon \left[ 3 - \left(\frac{r}{R} \right)^2 \left( f\sin^2\theta - 1 \right) \right] \nonumber
\ee
where we have defined the ratio of escape to thermal speed
\be
\lambda & =& \frac{GM_p}{R a^2}
\nonumber \\ & \simeq &
9.3 \left( \frac{M_p}{0.7M_J} \right) \left( \frac{1.4R_J}{R} \right)
\left( \frac{10\ \rm km\ s^{-1}}{a} \right)^2
\label{eq:lambda}
\ee
and the ratio of rotation speed (or tidal potential) to thermal speed
\be
\epsilon & = &  \left( \frac{\Omega R}{a} \right)^2
\nonumber \\ & = &
 0.043
\left( \frac{\rm 3.5\ days}{P_{\rm orb}} \right)^2
\left( \frac{R}{1.4R_J} \right)^2
\left( \frac{10\ \rm km\ s^{-1}}{a} \right)^2.
\label{eq:epsilon}
\ee
The photoionization model in section \ref{sec:Hlayer} shows
outward increase in $T$ and decrease in $\mu$, implying larger $a$
with radius. The density will then decrease more slowly than in the 
isothermal model for the same quantities at the base.

%We note that the density
%is larger using dipole geometry than using spherical geometry since
%$U(R,\theta_b,\phi)>U(R,\theta,\phi)$, due to the latitude dependence
%of the tidal force.

\begin{figure}[t]
\epsscale{1.2}
\plotone{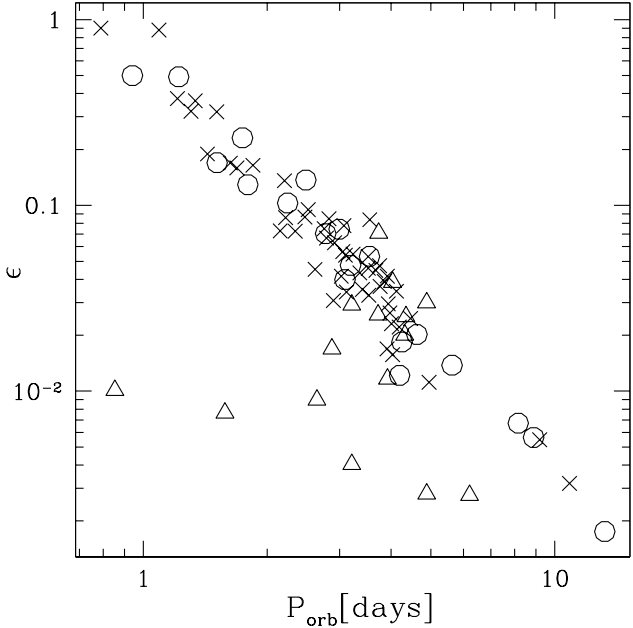}
\caption{
Values of $\epsilon$ calculated using eq.\ref{eq:epsilon} for the
current database of transiting exoplanets (http://exoplanet.eu/).
The open circles, X's and open triangles are planets with
$M_p/M_J>2$, $1/2 \leq M_p/M_J \leq 2$ and $M_p/M_J<1/2$, respectively.
The parameters used are $a=9.3\ {\rm km\ s^{-1}}$, appropriate
to $T=10^4\ {\rm K}$, $\mu=1$, and $R= R_{\rm ph}$.
}
\label{fig:epsobs}
\end{figure}

Figures \ref{fig:lambdaobs} and \ref{fig:epsobs} show the values of
$\lambda$ and $\epsilon$ versus planet orbital period for the transiting
exoplanets, taking $M_p$, transit radius $R_{\rm ph}$ and orbital period
$P_{\rm orb}=2\pi/\Omega$ from the Extrasolar Planets Encyclopedia\footnote{http://exoplanet.eu/}. The temperature and mean molecular weight have
been set to fiducial values of $\mu=1$ and temperature
$T=10^4\ {\rm K}$, giving sound speed of $a=9.3\ {\rm km\ s^{-1}}$.

Note that a substantial number of planets have $\lambda=2-10$, implying
the scale height of the gas is large enough that the density decrease
in the dead zone is only by a factor of $10-10^4$, far less than for
planets with cold upper atmospheres more distant from their parent
star. This increased density leads to the possibility that hot Jupiter
upper atmospheres may be collisional to large distances from the planet,
i.e. that the exobase, if it exists at all, is at radii $r \gg R_{\rm
ph}$ \citep{2005ApJ...621.1049T,2009ApJ...693...23M}.

Next, note that for the same fiducial molecular weight and temperature,
the strength of the tide, $\epsilon$, is in the range $0.1-1$ for
a substantial number of planets.  For large $\epsilon$, the typical
rotational speed of a synchronized planet is comparable to the
sound speed, or equivalently, the free fall speed in the tidal
potential is comparable to the sound speed.  Since $\epsilon$ is
evaluated at the base, the tidal force will dominate even
more at larger distances from the planet. 
%Hence, the acceleration
%of the wind in these planets is driven, to a large extent, by the
%tidal force, rather than being driven by gas pressure gradients.

\begin{figure}[t]
\epsscale{1.2}
\plotone{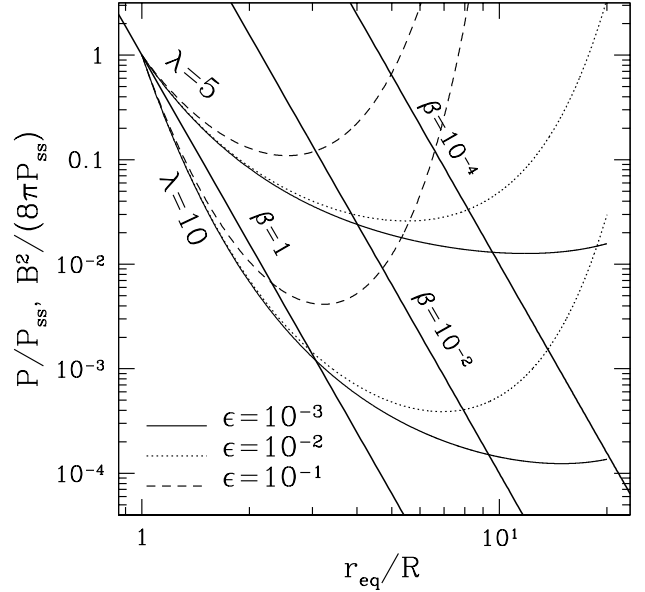}
\caption{ Gas and magnetic pressure, normalized to the base gas
pressure $P_{ss}=P(R,\pi/2,0)$, as a function of equatorial radius $r_{\rm
eq}$ for the isothermal model. The tidal potential is evaluated
along the star-planet line $\cos^2\phi=1$ at the equator
$\theta=\pi/2$. The two groups of lines
starting from $r=R$ and $P=P_{ss}$ are $P(r_{\rm eq},\pi/2,0)/P_{ss}$
evaluated for $\lambda=5,10$. For each group, the line style gives
the value of $\epsilon$. The three lines sloping down to the right
are $B^2(r_{\rm eq},\pi/2,0)/(8\pi P_{ss})=(R/r_{\rm eq})^6/\beta$ for
the three different values of equatorial $\beta=10^{-4},10^{-2},1$.
The cusp radius in Figure \ref{fig:cusp} is given by the intersection
of the gas and magnetic pressure curves.  The gas pressure decreases
outward faster for larger $\lambda$. Beyond the Roche radius, gravity
effectively points outward and the gas pressure begins to increase
outward. For larger $\epsilon$, the Roche radius moves inward.
}
\label{fig:pvsr}
\end{figure}

Figure \ref{fig:pvsr} shows the run of gas and magnetic pressure
along the equator ($\theta=\pi/2$) as a function of radius along
the star-planet line ($\phi=0$). Eq.\ref{eq:isoP} was used for the gas
pressure, and eq.\ref{eq:dipole} for the magnetic pressure. 
The magnetic pressure is parametrized by the plasma $\beta$
at the substellar point at the base:
\be
\beta & \equiv & \frac{ 8\pi P_{ss} }{(B_0/2)^2}
=
0.14\ \left( \frac{P_{ss}}{0.1\ \rm \mu bar} \right) \left( \frac{B_{\rm J,eq}}{B_0/2}
\right)^2.
\label{eq:beta}
\ee
The different lines in Figure \ref{fig:pvsr} show gas pressure
and magnetic pressure for different $\lambda$, $\epsilon$, and $\beta$
as a function of radius.

For small $\epsilon$, the density decreases outward, and eventually
becomes a constant. When the tidal force is included the density
{\it increases} outward for radii outside the magnetic Roche radius
(eq.\ref{eq:rRB}), since the sign of gravity points outward there.
This is a dramatic effect for close-in planets, whose Roche radii
are at only a few planetary radii, and may lead to hydrogen densities
orders of magnitude larger than the $\epsilon=0$ case. 
%In the region
%of outwardly increasing density, the gas is supported by magnetic
%forces. 
The tidal gravity plays a role similar to the centrifugal
force in models of the closed field line regions in the solar wind
\citep{1987MNRAS.226...57M} and in Jupiter's magnetosphere outside
the corotation radius. To specify the current and field distributions
required for this support involves a solution of the trans-field
equation, which is beyond the scope of this work. However, such
support should be possible when magnetic pressure dominates gas
pressure.

\begin{figure}[t]
\epsscale{1.2}
\plotone{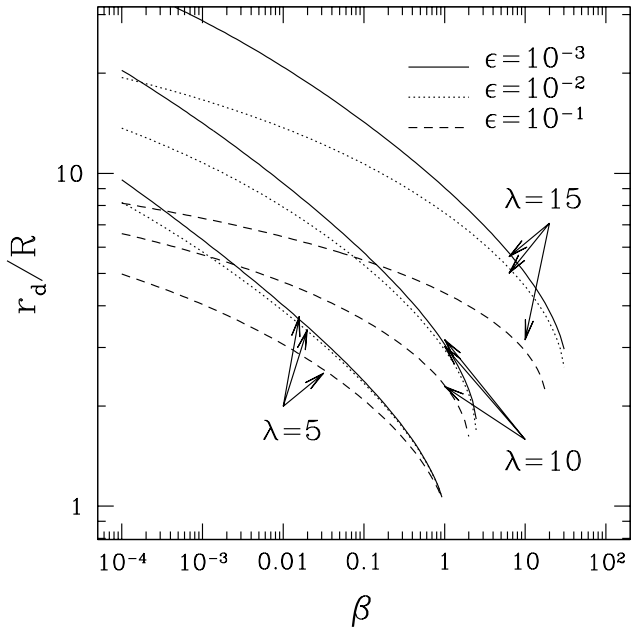}
\caption{
Cusp radius as a function of equatorial plasma $\beta$ at the substellar
base of the atmosphere for different values of $\lambda=GM_p/(R a^2)$
and $\epsilon=(\Omega R/a)^2$.  Here $\lambda$ is roughly the ratio of
escape speed to isothermal sound speed, and $\epsilon$ is the roughly
the ratio of rotation speed to isothermal sound speed.  Larger $\lambda$
($\epsilon$) implies the density decreases outward faster (slower).
The cusp radius moves outward (inward) for larger $\lambda$ ($\epsilon$).
}
\label{fig:cusp}
\end{figure}

Next, we follow \citet{1968MNRAS.138..359M} and
\citet{1987MNRAS.226...57M} to estimate the size of the dead and wind
zones.  \citet{1971SoPh...18..258P} showed that the dead zone ends at
the equator in a cusp, i.e. the field in the dead zone approaches zero
toward the cusp. Also,  for $v \ga a$, the wind zone pressure can be
neglected compared to the dead zone pressure, leading to the condition (also see our discussion leading up to eq.\ref{eq:balance} in Appendix A)
\be
P_{\rm dead} & \simeq & \frac{B_{\rm wind}^2}{8\pi}.
\label{eq:cusp1}
\ee
To determine the cusp radius, which we call $r_d$ for dead zone, we use
eq.\ref{eq:isoP} for the left hand side and eq.\ref{eq:dipole} for the
right hand side. The resulting equation is
\be
&& \beta \exp\left[ -\lambda\left(1-\frac{R}{r} \right)
+ \frac{3}{2} \epsilon \left( \frac{r^2}{R^2} \cos^2\phi - 1 \right) 
 \right]  =  \left( \frac{R}{r} \right)^{-6}
\label{eq:cusp2}
\ee
Eq.\ref{eq:cusp2} shows that the cusp radius will depend on $\phi$
in general due to the tidal potential. When tides can be ignored
($\epsilon \ll 1$), a reasonable approximation is $r_d/R \simeq
(e^{\lambda}/\beta)^{1/6}$.

Figure \ref{fig:cusp} shows cusp radii as a function of $\beta$
for different tidal strength ($\epsilon$) and binding parameter
($\lambda$). Consider the fiducial case with $\epsilon=10^{-3}$,
$\lambda=10$ and $\beta=10^{-2}$. Initially the magnetic pressure
is much larger than the gas pressure. The more rapid decrease of
magnetic pressure implies equality at the cusp radius, $r_d/R \simeq
10$. Increasing $\epsilon$, the gas pressure is larger and the cusp
radius moves inward. The cusp radius moves outward with increasing
magnetic field, i.e. decreasing $\beta$.

Under what conditions does a magnetosphere not form? Inspection of
the $\beta=1$ and $\lambda=5$ curves in Figure \ref{fig:pvsr} shows
that the magnetic and gas pressures are initially equal at the base,
but the gas density decreases more slowly and so the magnetic field
never dominates. Ignoring tides, there is an analytic criterion for the
critical $\beta_{\rm crit}=\beta_{\rm crit}(\lambda)$ for the formation
of a dead zone. This criterion is found by requiring simultaneously
$P=B^2/8\pi$ and $dP/dr=d/dr(B^2/8\pi)$, and yields $\beta \leq \beta_{\rm
crit}=(6/\lambda)^6\exp(\lambda-6)$. For $\lambda=5,10,15$, the values
are $\beta_{\rm crit}=1.1, 2.5$ and $33$, agreeing with the cutoffs in
Figure \ref{fig:cusp}. Hence for large $\lambda$, the gas at the base
need not be magnetically dominated in order for the gas well above to
base to become so.

In summary, we have found cusp, or dead zone, radii in the range of a
few to tens of radii $R$ for the expected range of $\lambda$, $\beta$
and $\epsilon$. Since transit observations to date probe the high density
gas within a few planetary radii, these observations may be probing
static gas trapped within the magnetosphere, as opposed to outflowing gas.

Nevertheless, as we will argue in section
\ref{sec:wind}, a wind zone should exist, and we investigate its
structure in the next section.

%%%%%%%%%%%%%%%%%%%%%%%%%%%%%%%%%%%%%%%%%%%%%%%%%%%%%%%%%%%%%%%%%%%%%

\section{ the wind zone }
\label{sec:wind}

%In this section we show how the (slow magneto-) sonic point is affected
%by the magnetic geometry. Readers unfamiliar with the MHD wind equations
%can consult Appendix \ref{sec:mhd} for a brief summary of the equations.
%We simplify the problem by assuming that the sonic point is close
%enough to the planet for magnetic stresses to dominate over hydrodynamic
%stresses -- the rigid field line approximation. In this situation the fluid nearly corotates with the planet (and can be accelerated like ``beads on a wire"). In
%the hot Jupiter case, the strong tidal forces can also contribute to
%acceleration by the combined ``magneto-centrifugal-tidal" effect from
%stellar wind theory \citep{1968MNRAS.138..359M}, which becomes important
%when the rotation velocity approaches the sound speed at the sonic point
%\citep{1987MNRAS.226...57M}. For a synchronized planet, the condition
%that the rotation velocity is comparable to the sound velocity at the
%sonic point is equivalent to the Roche-lobe radius being near the sonic
%point. The rigid field line assumption will typically break down
%near the Alfv\'en point, which we estimate lies well outside the dead zone
%radius for most latitudes, outside the region of interest near the planet.
In this section we show how the (slow magneto-) sonic point is affected by
the magnetic geometry. Readers unfamiliar with the MHD wind equations can
consult Appendix \ref{sec:mhd} for a brief summary of the equations. We
simplify the problem by assuming that the sonic point is close enough to
the planet for magnetic stresses to dominate over hydrodynamic stresses
--- the rigid field line approximation. In this situation the fluid
nearly corotates with the planet, and can be accelerated like ``beads on
a wire" by the magnetic field. This ``magneto-centrifugal" effect
from stellar wind theory \citep{1968MNRAS.138..359M} becomes important
when the rotation velocity approaches the sound speed at the sonic
point \citep{1987MNRAS.226...57M}. For a synchronized hot Jupiter, tidal
forces are of comparable size as centrifugal forces, and the condition
that the rotation velocity is comparable to the sound velocity at the
sonic point is equivalent to the Roche-lobe radius being near the sonic
point. The rigid field line assumption will typically break down near
the Alfv\'en point, which we estimate lies well outside the dead zone
radius for most latitudes.

Plugging parallel velocity, $\vec{v}=v\vec{b}$, and the no monopoles
condition, eq.\ref{eq:divB}, into eq.\ref{eq:cont}, the continuity
equation assumes the simple form
\be
\vec{B} \cdot \grad \left( \frac{\rho v}{B} \right) & = & 0
\label{eq:cont2}
\ee
so that $\rho v/B$ is constant on field lines, and has the interpretation
of the mass loss rate per unit of magnetic flux along a flux tube of
area $\propto B^{-1}$. Similarly, the projection of eq.\ref{eq:mom}
along the field can be rewritten as
\be
v \frac{dv}{ds}  & = & - a^2 \frac{d\ln \rho}{ds} - \frac{dU}{ds},
\label{eq:mom3}
\ee
which has the Bernoulli integral along field lines (see eq.\ref{eq:mom2}
and \ref{eq:bernoulli}). Again, the $(\vec{J} \times \vec{B})$
force cancels out since $\vec{v} \ \| \ \vec{B}$. Combining eq.\ref{eq:cont2}
and \ref{eq:mom3} gives the momentum equation parallel to field lines:
\be
\left( v - \frac{a^2}{v} \right) \frac{dv}{ds} & = & 
- a^2 \frac{d\ln B}{ds} - \frac{dU}{ds}.
\label{eq:criticalpoint}
\ee
At the critical point, $v=a$, and hence to avoid a divergent acceleration,
the right hand side must go to zero, giving
\be
- a^2 \frac{d\ln B}{ds} & = & \frac{dU}{ds}
\label{eq:nozzleeqn}
\ee
at the sonic point.  The term on the left hand side represents the
pressure gradient due to the geometry set by the magnetic field. The
term on the right hand side is the net acceleration along the field line.

Since the sonic point is sufficiently close to the surface that the
field geometry is not much perturbed by external currents, we use
dipole geometry to evaluate the sonic point position. The dipole field
in eq.\ref{eq:dipole} gives the intermediate result
\be
- \frac{d\ln B}{ds} & =& \frac{3\cos\theta}{Nr} \left( 1 + \frac{\sin^2\theta}{8N} \right).
\label{eq:dlnBds}
\ee
The two terms on the right hand side of eq.\ref{eq:dlnBds} represent
the field line divergence due to the $r^{-3}$ factor from dipole
field geometry, and the $\theta$-dependent factor $N$.  The term due
to differentiating $N$ is negligible for large loops, but becomes
important for small loops near the equator.  Plugging eq.\ref{eq:apar}
and \ref{eq:dlnBds} into eq.\ref{eq:nozzleeqn}, and eliminating $\theta$
using the field line geometry in eq.\ref{eq:fieldline}, we find the
following equation to determine the sonic point $r=r_s$:
\be
\frac{GM_p}{r^2} + \Omega^2 r & =& \frac{3a^2}{r}
%+ \frac{3a^2}{8Nr_{\rm eq}} + \frac{3f\Omega^2 r^2}{2r_{\rm eq}}
\label{eq:speqn}
\ee
where $N=\sqrt{1-3r/4r_{\rm eq}}$. This equation, solely in terms of
$r$, can be solved as a function of the parameters $\lambda$, $\epsilon$
and $r_{\rm eq}/R=1/\sin^2\theta_b$.

First we examine the limit in which tidal forces can be neglected. This
simple case highlights the importance of the magnetic field geometry,
and would apply for slow rotating planets distant from the star.
Setting $\Omega=0$ in eq.\ref{eq:speqn}, the simpler equation
\be
\frac{GM_p}{3a^2r} & =&
1 + \frac{r}{8r_{\rm eq}\sqrt{1-3r/4r_{\rm eq}}}
\label{eq:speqn2}
\ee
results. For large field lines $r_{\rm eq} \gg R$, field line curvature
is negligible and the sonic point sits at
\be
r_{s0} & \simeq & \frac{ GM_p}{3a^2}
= \left( \frac{\lambda}{3} \right) R
\label{eq:rs0}
\ee
which differs from the spherical wind result by the 3, instead of 2,
in the denominator. Including finite $\sin^2\theta_b$, the sonic point
moves inward somewhat.

%In the unmagnetized case,
%sonic point solutions for a spherical wind may always be found. In the
%magnetized case, for flow along dipole field lines, solutions for the
%sonic point are not possible when the loop is smaller than a critical
%size $r_{\rm eq}=(4/5)r_{s0}$ at which point the left hand side is always
%larger than the right hand side in eq.\ref{eq:speqn2}. In other words,
%no sonic point solution is possible when the curvature radius of the
%field line becomes smaller than the zero field sonic point radius in
%eq.1\ref{eq:rs0}.

%One could ask the question if it is possible that
%there are no solutions to eq.\ref{eq:speqn2} in the wind zone, so that our assumption
%of a equatorial dead zone and polar wind zone are violated. We find that indeed
%if $r_{\rm eq} \la r_{s0}$, there is no solution to eq.\ref{eq:speqn2} for the sonic point, and no wind
%can be driven on that field line. However, for realistic parameters, we find that
%field lines outside the dead zone will indeed have solutions for the sonic point. The reason
%is that $r_d/R \simeq e^{\lambda/6}/\beta^{1/6}$ increases quite rapidly with $\lambda$.
%One can then show that to shut down field lines in the wind zone, you would need $\beta \gg 1$
%at the base of the flow, contradicting our assumption large magnetic field $\beta \ll 1$.

\begin{figure}[t]
\epsscale{1.2}
\plotone{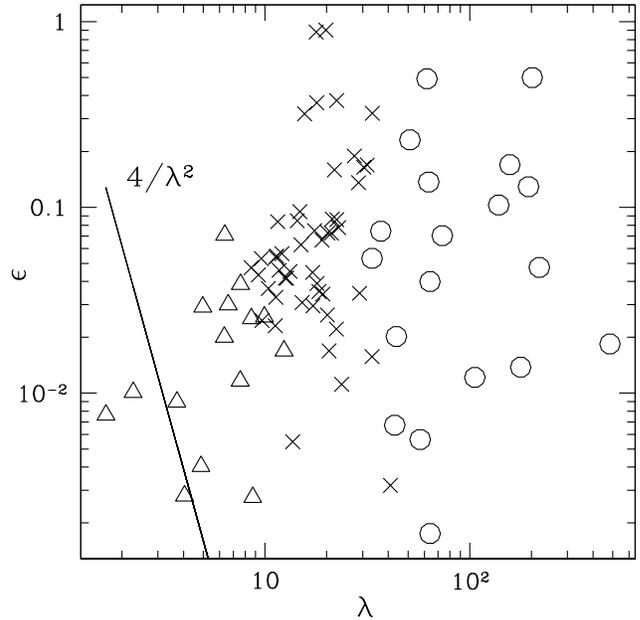}
\caption{ 
Values of $\epsilon$ versus $\lambda$ for the transiting planets. The
points are the data, with symbol type as in Figures \ref{fig:lambdaobs}
and \ref{fig:epsobs}. The line is the critical tidal strength
$\epsilon=4/\lambda^2$ above which the wind is suppressed in the polar
region.  We have used $a=9.3\ {\rm km\ s^{-1}}$ to make the plot.
}
\label{fig:eps_vs_lam}
\end{figure}

Next, we include tidal forces, but ignore the field line curvature
terms on the right hand side scaling as $r_{\rm eq}^{-1}$, a good
approximation for field lines near the pole.  This approximation
eliminates the possibility that the tidal force can point outward.
In this case, the sonic point equation becomes
\be
\frac{GM_p}{r^2} + \Omega^2 r & =& \frac{3a^2}{r}.
\label{eq:speqn3}
\ee
The key point is that now the effective gravity on the left hand side
has a {\it minimum}.  If the pressure term on the right hand side
is smaller than this minimum, a transonic solution is not possible.
The solution of the cubic eq.\ref{eq:speqn3} lying near the planet
disappears for sufficiently strong tidal forces
\be
\Omega & \geq & \Omega_{\rm crit} = 2 \frac{a^3}{GM_p} = \frac{2}{3} \frac{a}{r_{s0}}
\nonumber \\ & = & 
 \left(\frac{2\pi}{3.4\ \rm days} \right) \left( \frac{0.7\ M_J}{M_p} \right)
\left( \frac{a}{10\ \rm km\ s^{-1} } \right)^{3}.
\label{eq:Omegacrit}
\ee
%at which point the sonic point has moved outward to $r_s=GM_p/2a^2$.
Eq.\ref{eq:Omegacrit} can also be written in dimensionless form as 
\be
\epsilon_{\rm crit} & = & \frac{4}{\lambda^2}.
\label{eq:epscrit}
\ee
When eq.\ref{eq:Omegacrit} is satisfied, for planets sufficiently close
to the star, there is no sonic point solution, and the wind is shut off
near the poles.  The density distribution on these field lines will be
hydrostatic.  Hence, for sufficiently strong tides such that the rotation
velocity $\Omega_{\rm crit} r_{s0}$ at the sonic point is supersonic,
a second dead zone is created in the polar regions where tides point
inward. Whereas the first dead zone in the equatorial region is due to
magnetic pressure dominating gas pressure, the second dead zone near
the poles arises due to the large potential barrier.

Figure \ref{fig:eps_vs_lam} shows $\epsilon$ versus $\lambda$ for the
observed transiting planets. Except for a handful of planets with the
smallest values of both $\epsilon$ and $\lambda$, most of the planets
are in the strong tide limit with $\epsilon > \epsilon_{\rm crit}$.
The planets in the upper right hand corner will have the polar wind
partially shut off, while the planets in the lower left hand corner
will be able to drive a polar wind.

Next, we retain the $r_{\rm eq}^{-1}$ terms due to the tidal force in
eq.\ref{eq:speqn}.  Taking the limit $\Omega \rightarrow \infty$ in
eq.\ref{eq:speqn}, the sonic point in the strong tide limit is
\be
r_s & \rightarrow & \frac{2r_{\rm eq}}{3f}.
\label{eq:rsrhb}
\ee
In this limit, the sonic point occurs at a fixed fraction of the loop
equatorial radius, dependent only on $\cos^2\phi$, and agrees with the
magnetic Roche lobe radius found in eq.\ref{eq:rRB}, where the net force
first points outward.

\begin{figure}[t]
\epsscale{1.2}
\plotone{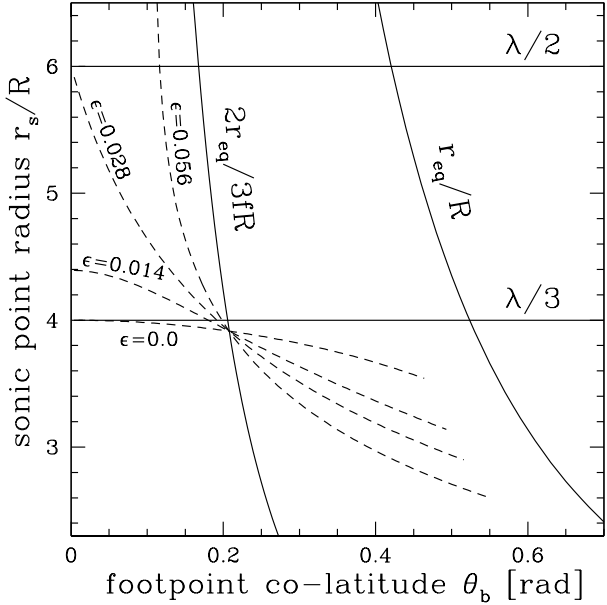}
\caption{ 
Sonic point radius as a function of footpoint co-latitude
for $\lambda=12$, $\beta=1$ and a range of $\epsilon$ (dashed
lines).  The three solid lines label the $\epsilon=0$ sonic point,
$r_{s0}/R=\lambda/3$, the $\epsilon=4/\lambda^2=0.16$ sonic point,
$r_s/R=\lambda/2$, and the $\epsilon=\infty$ sonic point, $r_s=2r_{\rm
eq}/3f$.  Each dashed line terminates at large $\theta$ at the dead
zone, $\theta=\theta_d$.  The dead zone shrinks ($\theta_d \rightarrow
\pi/2$) as $\epsilon$ increases.  The value of $\beta$ is needed only
for the size of the dead zone.  The longitude $\cos^2\phi=1$, along the
star-planet line, has been assumed here.
}
\label{fig:sonic}
\end{figure}

Figure \ref{fig:sonic} shows an example numerical solution of
eq.\ref{eq:speqn} for the sonic point radius as a function of footpoint
angle $\theta_b$. Dipole geometry was used to produce this plot.
In the weak tide limit ($\epsilon \rightarrow 0$),
the solutions asymptote to eq.\ref{eq:rs0} for large loop size, and
decrease slightly before terminating at the dead zone, $\theta=\theta_d$.
For small $\epsilon \la 4/\lambda^2$, the sonic point moves out in
the polar regions, and inward closer to the equator; the dividing line
between these two behaviors depends on if the net force is outward or
inward near the looptop.
% Since the mass loss rate is proportional to
%the density at the sonic point, there will be larger mass loss rate for
%field lines nearer the equator than the poles. 
Next, for the critical
value $\epsilon=4/\lambda^2$, the sonic point is at roughly $r_s \simeq
(\lambda/2)R$ near the pole. For larger values $\epsilon \ga 4/\lambda^2$,
the sonic point near the pole jumps out to a radius much further from the
planet, near $r_s \sim 2r_{\rm eq}/3f$, where the net gravity changes
sign. For small values of $\theta_b$, this solution may be at tens to
hundreds of planetary radii, and is of no physical interest. Physically,
when the sonic point moves outside the region of interest the field line
is effectively hydrostatic.

\begin{figure}[t]
\epsscale{1.2}
\plotone{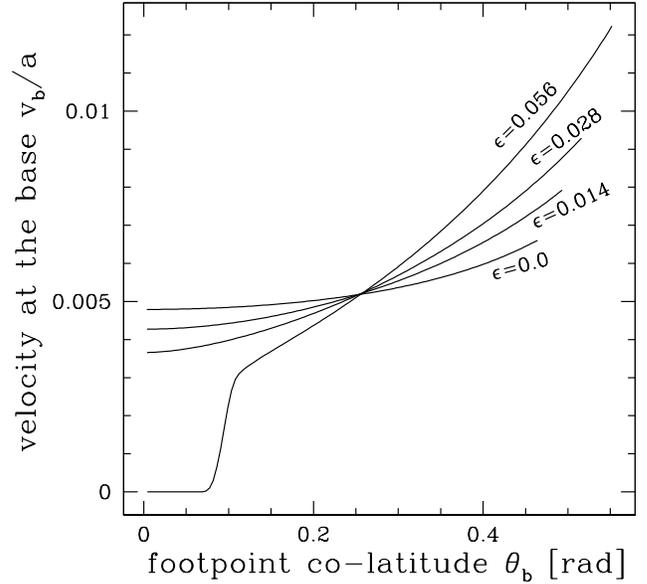}
\caption{ 
Velocity at the base, in units of $a$, for the same parameters as Figure
\ref{fig:sonic}.
}
\label{fig:vbase}
\end{figure}

Figure \ref{fig:vbase} shows the velocity at the base, $v_b$, for
the same parameters as in Figure \ref{fig:sonic}. This velocity was
found by using the Bernoulli constant evaluated at the sonic point
and the base. For weak tides ($\epsilon=0$), the velocity at the
base is nearly constant over the wind zone.  As $\epsilon$ increases,
the base velocity decreases near the pole, where the sonic point
has moved outward, and increases closer to the equator, where the
sonic point has moved inward. 

At sufficiently large radii, pressure gradients rapidly become negligible
and the fluid should move on a nearly ballistic trajectory. Using the
Bernoulli integral defined in eq.\ref{eq:bernoulli}, with the
sonic point as reference location, the velocity parallel to the magnetic
field is
\be
v^2 & = & a^2 + 2a^2 \ln \left( \frac{B_s}{B} \frac{v}{a} \right)
+ 2 \left( U_s - U \right)
\label{eq:eqforv}
\ee
where the subscript ``s" refers to the sonic point. The transit signal
depends only on the gas velocity within the stellar disk at $r \leq
R_\star$, where $R_\star$ is the stellar radius.  The tidal potential term
$-2U$ dominates at large radius.  For the purposes of a simple estimate,
this asymptotic expression, evaluated at the stellar radius, gives
\be
v_{\rm asymp} & \simeq & \Omega R_\star \left( f \sin^2 \theta - 1 \right)^{1/2}
\nonumber \\ &  = & 
24\ {\rm km\ s^{-1}}\ \left( \frac{3.6\ \rm day}{P_{\rm orb}} \right)
\left( \frac{R_\star}{R_\odot} \right) \left( \frac{f \sin^2 \theta - 1}{3} \right)^{1/2}.
\label{eq:vasymp}
\ee
We caution the reader that the (logarithmic) enthalpy term in eq.\ref{eq:eqforv} is
not negligible, and can increase the velocity at $r = R_\star$ by a
factor of order 2 (see Figure \ref{fig:fieldline}). Though supersonic,
the velocity in eq.\ref{eq:vasymp} is much smaller than the $\sim \pm 100\
{\rm km\ s^{-1}}$ at which absorption is observed in the Lyman $\alpha$
spectrum of HD 209458b \citep{2003Natur.422..143V}, and hence velocity 
gradients cannot be the origin of the observed transit depth.

%However, some results are
%sensitive to the field geometry. For instance, for wind directed up from
%the orbital plane, the assumption that fluid follows a straight field
%line is not valid, due to the high potential barrier. More likely is that
%the streamilines would bend back down toward the orbital plane, once the
%fluid ram pressure dominates magnetic pressure. Our straight line model
%will, we believe, artificially suppress the wind on such field lines.
%For the purposes of producing a column density map, this approximation is
%valid as the fluid would be nearly hydrostatic in the region of interest.

In the following sections, we will implement the general theory that
describes the structure of the dead zone (\S~\ref{sec:dead}) and of the
wind zone (\S~\ref{sec:wind}) to construct global models of hot Jupiter
magnetospheres for comparison with transit observations of HD 209458b.
A simplified 1D thermal model motivates our choice of the pressure and
sound speed at the base of the global models, which are key parameters
for determining the magnetospheric structure.

%%%%%%%%%%%%%%%%%%%%%%%%%%%%%%%%%%%%%%%%%%%%%%%%%%%%%%%%%%%%%%%%%%%%%

\section{ the H and H$^+$ layers: a simplified 1D thermal model }
\label{sec:Hlayer}

The thickness of the warm H layer with temperature $T \simeq 5,000-10,000\
{\rm K}$ is a crucial parameter in determining the transit depth, as
emphasized in the phenomenological model of \citet{2010arXiv1004.1396K} (see \citet{2007P&SS...55.1426G} for a discussion of the role
of base pressure for an atmosphere undergoing energy-limited escape).
The large temperature and small mean molecular weight increase the scale
height to $\sim 0.1 R_{\rm ph}$. If the warm layer extends over $\ga 10$
scale heights, the transit radius can be significantly increased over
the photospheric radius of the optical continuum.  % Reasonable values 
%for the thickness and base pressure of this model are needed as input
% to the 3D isothermal model in this paper.  
In this section, we 
% go beyond the isothermal approximation to 
construct a simple 1D model in photoionization and thermal 
equilibrium to determine the depth of the
warm H layer.

At sufficiently low density and temperature, the rates of collisional
ionization and 3-body recombination are slow compared to photoionization
and radiative recombination, respectively, and the ionization state
is set by a balance of the latter two processes.  We consider a pure
hydrogen gas for simplicity.  Let $n_e$, $n_p$ and $n_H$ be the density
of electrons, protons and hydrogen atoms.  Charge neutrality implies
$n_e=n_p$.  The ionization fraction in the ``on the spot" approximation
is found by solving the algebraic equation \citep{2006agna.book.....O}
\be
n_H J(N_H) & = & \alpha_B(T) n_e n_p = \alpha_B(T) n_p^2.
\label{eq:ioneq}
\ee
At the 50\% ionization point,
\be
n_H & = & n_p \equiv n_{\rm eq}   =   \frac{J}{\alpha_B}.
\ee
Here $\alpha_B(T) \simeq
2.6 \times 10^{-13}\ {\rm cm^3\ s^{-1}} (10^4\ {\rm K}/T)^{0.8}$ is
the case B radiative recombination rate \citep{2006agna.book.....O},
\be
J(N_H) & = & \int_{\nu_0}^\infty d\nu\ \varphi_\nu\ e^{-N_H\sigma_{\rm pi}(\nu)}
\sigma_{\rm pi}(\nu)
\label{eq:J}
%\simeq \frac{J_0}{1 + (\sigma_{\rm pi} N_H)^{1.5}}
%\label{eq:Jfit}
% \left( 6\ {\rm hr} \right)^{-1} \left( \frac{0.05\ {\rm AU}}{D} \right)^2,
%\nonumber \\ & \times &
%\left( \frac{1}{1 + (N_H \times 6 \times 10^{-18}\ {\rm cm^2})^{1.5}} \right)
%\label{eq:Jfit}
\ee
is the photoionization rate per H atom, $\varphi_\nu$ is the photon flux
per unit frequency interval, $N_H$ is the atomic hydrogen column from
the point in question to the star, $\sigma_{\rm pi}(\nu)=\sigma_{\rm
pi}(\nu_0/\nu)^3$ is the H atom bound-free cross section, the threshold
cross section is $\sigma_{\rm pi} =6.3 \times 10^{-18}\ {\rm cm^2}$, and
the threshold frequency is $\nu_0=13.6\ \rm eV/h$. The exponential factor
in eq.\ref{eq:J} takes into account attenuation of the stellar radiation.
Eq.\ref{eq:J} may be computed as a function of $N_H$, as described in
\citet{2006agna.book.....O}.

For the thermal balance, the dominant processes are photoelectric heating
from the ionization of hydrogen atoms \citep{2004Icar..170..167Y}, and cooling
by collisionally excited Lyman $\alpha$ emission from electron impacts
\citep{2009ApJ...693...23M}. Assuming 100\% efficiency of turning
photoelectron energy into heat, the heating rate per reaction is
\be
Q(N_H) & = & \int_{\nu_0}^\infty d\nu\ \varphi_\nu\ e^{-N_H\sigma_{\rm pi}(\nu)}\
\sigma_{\rm pi}(\nu)\ h(\nu-\nu_0).
%\nonumber \\ & \simeq  &
%\frac{Q_0}{(1 + 0.1\sigma_{\rm pi}N_H)^{1.5}}
\label{eq}
\ee
%where
%\be 
%Q_0 & = & 2.0 \times 10^{-16}\ {\rm erg\ s^{-1}}
%\left( \frac{0.05\ {\rm AU}}{D} \right)^2
%\label{eq:Q0}
%\ee
%is the optically thin rate, giving a mean photoelectron energy
%$Q_0/J_0 \simeq 2.7\ {\rm eV}$.
Balancing photoelectric heating and cooling by collisionally
excited Lyman $\alpha$ emission implies
\be 
n_H Q(N_H) & = & \Lambda(T) n_e n_H.
\label{eq:thermaleq}
\ee
Note that $n_H$ cancels out of eq.\ref{eq:thermaleq}.  The line cooling
coefficient for Lyman $\alpha$ is $\Lambda(T)=2.9 \times 10^{-19}\ {\rm
erg\ cm^3\ s^{-1}}\ \sqrt{10^4\ {\rm K}/T} \exp(-118,400\ {\rm K}/T)$
\citep{1972ARA&A..10..375D}. Increasing distance from the star and 
decreased heating efficiency act to decrease $Q$.
%If the EUV flux were smaller in magnitude,
%or the efficiency was less than 100\%, but
%with the same spectral shape, $Q$ could be decreased proportionally.

Eq. \ref{eq:ioneq} and \ref{eq:thermaleq} are two algebraic equations
which can be solved for $n_H$ and $n_p$ in terms of $N_H$ and
$P=(2n_p+n_H)k_bT$. In practice, we assume a trial $T$, compute $n_H$ and
$n_p$ from eq.\ref{eq:ioneq}, and then compute the imbalance of heating
and cooling in eq.\ref{eq:thermaleq}. The temperature is iterated until
thermal balance is achieved. The equations for dependent variables $N_H$ and
$P$ and independent variable $r$ are the definition of column
\be
\frac{dN_H}{dr} & = & - n_H
\label{eq:dNHdr}
\ee
and hydrostatic balance
\be
\frac{dP}{dr} & = & - \frac{GM_p m_p}{r^2} \left( n_H+n_p \right),
\label{eq:dPdr}
\ee
where tides have been ignored for simplicity. The inner boundary condition
is $r=R$ at the chosen base pressure. The column $N_H$ should go to zero
at the outer boundary. We enforce this boundary condition at the finite,
but large, radius $r=40R$.

\begin{figure}[t]
%\epsscale{1.1}
\epsscale{1.2}
\plotone{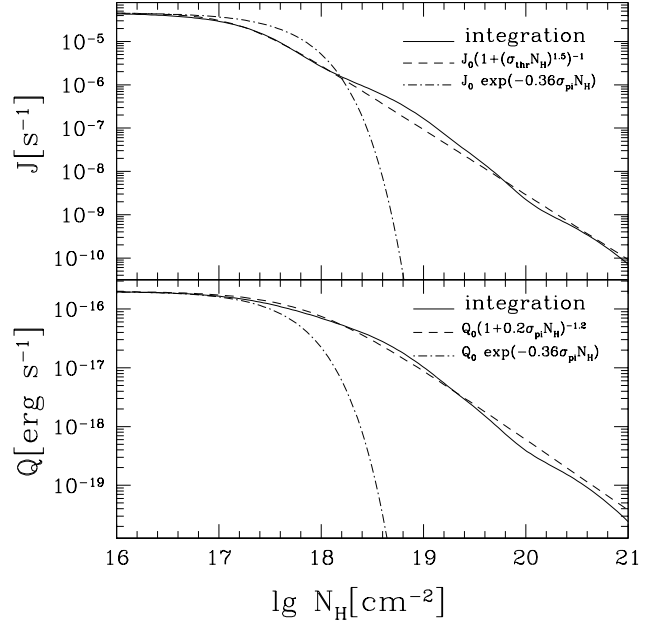}
\caption{ Photoionization rate (top) panel and heating rate (bottom)
as a function of neutral hydrogen column, for a planet at $D=0.05\ {\rm AU}$
around a solar-type star..
}
\label{fig:pirate}
\end{figure}

To compute the integrals in eq.\ref{eq:J} and eq.\ref{eq}, we use the quiet solar Lyman continuum spectrum from
\citet{1998SoPh..177..133W}, which tabulates $\int d\nu \varphi_\nu$
in each frequency bin. The results are shown as the solid lines in Figure \ref{fig:pirate}. Approximate power-law fits are also shown, along with curves representing pure exponential attenuation for comparison.
%\be
%J(N_H) & \simeq &  \frac{J_0}{1 + (\sigma_{\rm pi} N_H)^{1.5}} 
%\label{eq:Jfit}
%\ee
%\be 
%J_0 & \simeq & \left( 6\ {\rm hr} \right)^{-1} \left( \frac{0.05\ {\rm AU}}{D} \right)^2.
%\label{eq:J0}
%\ee
A similar fit with shallower slope is shown in the bottom panel for the
heating rate $Q(N_H)$. At small optical depth, $J_0 \approx 6$ hr and the mean photoelectron
energy is $Q_0/J_0 \simeq 2.7\ {\rm eV}$.  Given the form of the integrand
in eq.\ref{eq:J}, one might have expected an exponential scaling of
the form $J \propto \exp[-(\rm constant)\sigma_{\rm pi}N_H]$, implying
negligible heating and ionization deep in the H layer. This is shown
as the dot-dashed line in Figure \ref{fig:pirate}, and cuts off much
too sharply. The numerical result is better fit with a power-law, $J
\propto N_H^{-1.5}$, leading to larger heating rate deep into the H layer.

The weak scaling of $J$ with $N_H$ can be explained as the competition
between $\exp[-N_H\sigma_{\rm pi}(\nu)]$, which increases to higher
frequencies, and $\varphi_\nu$, which on average decreases to higher
frequency. Ignoring lines in $\varphi_\nu$, the product of these two
functions is a sharp peak, similar to the Gamow peak in thermonuclear
reaction rates. For instance, if one approximates the Lyman continuum as a
blackbody with temperature $T=8300\ {\rm K}$ \citep{1970SoPh...15..120N},
steepest descent evaluation of eq.\ref{eq:J} gives the weak exponential
scaling $J \propto \exp[ - 15.9 (N_H\sigma_{\rm pi})^{1/4} ]$, which
involves $N_H^{1/4}$ in the exponent rather than $N_H$.  However, the
blackbody fit is not adequate, as even this weaker scaling cuts off
too fast.  We find a better fit is to use a power-law form $\varphi_\nu
\propto \nu^{-\gamma}$, leading to $J \propto N_H^{-(\gamma+2)/3}$,
with no exponential scaling.  Choosing $\gamma=2.5$ then recovers the
observed scaling. While these analytic scalings are useful for intuition,
the solar Lyman continuum contains many strong lines, and is not well
approximated by a smooth continuum function when computing $J$.

\begin{figure}[t]
%\epsscale{1.1}
\epsscale{1.2}
\plotone{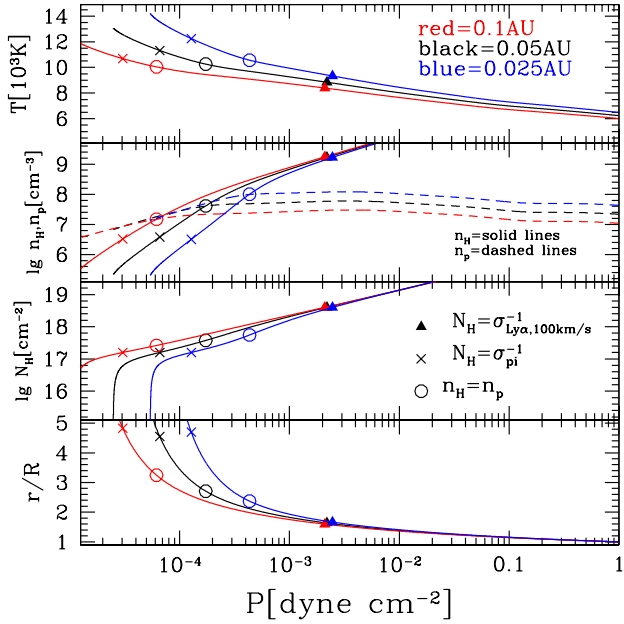}
\caption{ 
Temperature (top panel), number densities (second panel), neutral hydrogen
column (third panel), and radius (bottom panel) versus pressure for a
$M_p=0.7 M_J$, $R=1.3 R_J$ planet. The red, black and blue lines are
for planets at $D=0.1,0.05,0.025$ AU, respectively.  The hollow circles
show the position where Lyman continuum radiation (at threshold) from
the star becomes optically thick. The crosses show the 50\% ionization
point. The filled triangles show the position in the atmosphere where
stellar Lyman $\alpha$ photons at $\pm 100\ {\rm km\ s^{-1}}$ from line
center become optically thick.
}
\label{fig:pieq}
\end{figure}

Figure \ref{fig:pieq} shows a numerical integration of
eqs. \ref{eq:dNHdr}, \ref{eq:dPdr}, \ref{eq:ioneq} and
\ref{eq:thermaleq}. Spherically symmetric irradiation is assumed,
the base pressure is chosen to be $P=1\ {\rm dyne\ cm^{-2}}$, and
the $N_H \rightarrow 0$ condition is applied at the (arbitrary) radius
$r=40R_{\rm ph}$. Parameters appropriate for HD 209458b have been chosen
with $M_p=0.7M_J$ and $R=1.3R_J$. In each panel, the three different
color curves are for different semi-major axis; the semi-major axis of
HD 209458b is $D \simeq 0.05\ {\rm AU}$. In the top panel, note that $T$
increases slowly as the planet is moved closer to the star. A factor 16
increase in stellar flux translated into only a  10-20\% increase in
the pressure range of interest, due to the exponential $T$ dependence
in the cooling rate. Eq.\ref{eq:thermaleq} can be solved analytically as
\be
T & \simeq & \frac{1.1 \times 10^4\ {\rm K}}{ 1 + 0.089\ln[(n_e/n_{\rm eq,0})(Q_0/Q)]}.
\label{eq:Tsoln}
\ee
This simple estimate shows that the temperature is always near $10^4\
{\rm K}$ near the base of the photoionized layer where $n_H=n_p \simeq
n_{\rm eq}$.  As $n_e$ decreases below $n_{\rm eq}$, the temperature
rises logarithmically slowly.  This temperature rise toward small density
will increase the scale height and density at larger radii relative to
an isothermal model with the same base conditions.

The second panel shows the rapid inward increase of $n_H$, with a change
in slope at the H-H$^+$ boundary. The slow decrease of $n_p$ into the
atmosphere is due to the slow decrease of $J$ with $N_H$. If we had used
the exponential scaling $J \propto \exp(-N_H \sigma_{\rm pi})$ shown
in Figure \ref{fig:pirate}, $T$ and $n_p$ would have decreased inward
much more rapidly. The hollow circles show the position of $n_H=n_p$
at each $D$. Planets further from the star remain neutral higher up in
the atmosphere.

The third panel shows $N_H$, and x's show the position of $N_H \sigma_{\rm
pi}=1$, where the Lyman continuum at threshold is optically thick. The
optical depth unity point moves to lower pressure for planets more distant
from the star, although the radius at this point is relatively constant.

The bottom panel shows radius in units of base radius. The warm H
layer with temperature $T \simeq 10,000\ {\rm K}$ at pressures $P=\rm
(0.1-100)\ nbar$ contributes an amount $\simeq (1-2) \times R_{\rm ph}$
to the radius. Specifically, even the region below $N_H\sigma_{\rm
pi}=1$ can be warm enough to contribute significantly to the radius. In
our reference global model below, we use a base pressure $P_{\rm ss} =
40$ nbar (see Model 1 of Table~1).  

We now discuss how the simple 1D model differs from previous
investigations. One crucial difference is that the dead zone
should be hotter than the wind zone, for which adiabatic expansion is an
important coolant. \citet{2004Icar..170..167Y} included heating arising
from photoionization of hydrogen, but ignored the attenuation of the
stellar Lyman continuum into the atmosphere, leading to an overestimate of
the heating rate. This attenuation of EUV in the H layer will also lead to
smaller heating by H$_2$ photoionization much deeper in the atmosphere.
\citet{2009ApJ...693...23M} included finite optical depth, but enforced
an exponential cutoff which led to a steep temperature drop below the
$N_H\sigma_{\rm pi}=1$ point. As shown in Figure \ref{fig:pirate}, the
exponential cutoff is too rapid, and the slower power-law cutoff found
here gives additional heating deeper in the atmosphere.

Figure \ref{fig:pieq} shows that the $n_H=n_p$ and $N_H\sigma_{\rm pi}=1$
points occur near each other at $D=0.05\ {\rm AU}$, hence attenuation due
to finite optical depth cannot be ignored.  The relative position of the
$n_H=n_p$ and $N_H\sigma_{\rm pi}=1$ layers can be estimated as follows.
The number density at which $n_p=n_H$ is
\be 
n_{\rm eq} & = & \frac{J}{\alpha_B}
\nonumber \\ & \simeq & 
1.8 \times 10^8\ {\rm cm^{-3}} \left( \frac{T}{ 10^4\ {\rm K} } \right)^{0.8}
\left( \frac{0.05\ {\rm AU}}{D} \right)^2 \left( \frac{J}{J_0} \right).
\label{eq:neq}
\ee
This can be converted into a pressure as
\be
P_{\rm eq} & =&  3 k_b T n_{\rm eq}
\nonumber \\ & =&
 8.0 \times 10^{-4}\ {\rm dyne\ cm^{-2}} \left( \frac{M_p}{M_J} \right)
\left( \frac{10^{10}\ {\rm cm}}{R} \right) \left( \frac{10}{\lambda} \right)
\nonumber \\ & \times &
\left( \frac{0.05\ {\rm au}}{D} \right)^2 \left( \frac{T}{10^4\ {\rm K}} \right)^{0.8}
\left( \frac{J}{J_0} \right)
\label{eq:Peq}
\ee
The $D^{-2}$ scaling implies that the atmosphere becomes neutral out
to smaller pressures as the planet moves away from the star.
To estimate where $N_H \sigma_{\rm pi}=1$, we assume $n_p \gg n_H$, 
$P \simeq 2k_bTn_p$, and eq.\ref{eq:ioneq} gives $n_H \simeq n_p^2/n_{\rm eq,0}$,
where $n_{\rm eq,0}=J_0/\alpha_B$. For H atom scale height $\simeq k_bT/m_p g$, 
the pressure at which $N_H\sigma_{\rm pi}=1$ is
\be
P_{\rm N_H\sigma_{\rm pi}=1} & \simeq & 
2k_b T \left( \frac{ n_{\rm eq,0} m_p g }{k_b T \sigma_{\rm pi} } \right)^{1/2} \propto D^{-1}.
\label{eq:Ptaueq1}
\ee
The scaling with $D$ in eq.\ref{eq:Ptaueq1} implies that planets further
from the star will become optically thick at lower pressure. Comparing
the scalings in eq.\ref{eq:Peq} and \ref{eq:Ptaueq1}, one may expect
50\% ionization to occur outside $N_H\sigma_{\rm pi}=1$ at sufficiently
large $D$.

Lastly, we note that since the cooling rate $\propto n_e n_H$, we may
expect Lyman $\alpha$ cooling to be inefficient at both high and low
density, where other cooling mechanisms, such as heat conduction, may
be more efficient.

%%%%%%%%%%%%%%%%%%%%%%%%%%%%%%%%%%%%%%%%%%%%%%%%%%%%%%%%%%%%%%%%%%%%%

\section{ global models }
\label{sec:global}

We now attempt to construct global models for the magnetosphere, in 
order to compute the planetary transmission spectrum and mass loss 
rates.  We assume the following magnetic field model
\citep{1968MNRAS.138..359M,1974MNRAS.166..683O}:
\be
\vec{B}(r,\theta) & = & 
\left\{ 
\begin{array}{c}
B_0 \left( \frac{R}{r} \right)^3 \left( \vec{e}_r \cos\theta + \frac{1}{2} \vec{e}_\theta
\sin\theta \right), \ \ \ (r<r_{\rm d}) \\
B_0 \left( \frac{R}{r_{\rm d}} \right)^3 \left( \frac{r_{\rm d}}{r} \right)^2 \cos\theta
\vec{e}_r, \ \ \ (r>r_{\rm d}).
\end{array}
\right.
\label{eq:Bmodel}
\ee
This global field model allows a position $(r,\theta)$ to be associated
with a base co-latitude $\theta_b$ at $r=R$.  The field is dipole for
$r<r_d$, and radial outside $r_d$. This radial field is distinct from
the split monopole due to the $\cos\theta$ factor.  Eq.\ref{eq:Bmodel}
is approximately correct for the dead zone, and also for determining
the sonic point position in the wind zone if the sonic point is close
to the planet.  However, as ballistic trajectories with speeds far less
than escape speed are expected to be bent down toward the orbital plane,
the radial field assumption will be unphysical for some latitudes.

Given parameters $\lambda$, $\epsilon$, $\beta$, and
longitude $\phi$, we first solve eq.\ref{eq:cusp2} for $r_d$
using dipole geometry. For field lines inside the dead zone,
$\sin\theta_b>\sin\theta_d=\sqrt{R/r_d}$, the velocity is zero and the
density is given by the hydrostatic expression in eq.\ref{eq:isoP},
where $P_{\rm ss}$ is a model parameter.  On field lines outside the
dead zone, $\sin\theta_b < \sin\theta_d=\sqrt{R/r_d}$, we search for
sonic points by looking for minima of the quantity $-\ln(B/B_0) - U/a^2$
on field lines, between the base radius $R$ and a chosen maximum radius
$r_{\rm max}$. Given the position of the sonic point $(r_s,\theta_s)$, the
Bernoulli equation $v^2/2 + a^2\ln(B/v) + U= {\rm constant}$ can be used
to solve for the velocity at the base, $v_b(\theta_b,\phi)$.  Given $v_b$
and the base density (eq.\ref{eq:rhob}), the Bernoulli equation may
again to used to find the run of $v$ and $\rho$ on the field line.

\begin{deluxetable*}{p{0.7cm}cccccccccccccccc}
%\tabletypesize{\scriptsize}
\tablecolumns{17}
\tablewidth{0pt}
\tablecaption{ Global Models \label{tab:models} }
\tablehead{
\colhead{model} &
\colhead{$M_p$\tablenotemark{a}} &
\colhead{$R_{\rm ph}$\tablenotemark{b}} &
\colhead{$D$\tablenotemark{c}} &
\colhead{$R$\tablenotemark{d}} &
\colhead{$P_{\rm ss}$\tablenotemark{e}} &
\colhead{$a $\tablenotemark{f}} & 
\colhead{$B_0 $\tablenotemark{g}} &
\colhead{$\lambda$\tablenotemark{h}} &
\colhead{$\epsilon$\tablenotemark{i}} &
\colhead{$\beta$\tablenotemark{j}} &
\colhead{$r_L $\tablenotemark{k}} &
\colhead{$r_{s,0}$\tablenotemark{l}} &
\colhead{$r_d$\tablenotemark{m}} &
\colhead{$\delta F/F$\tablenotemark{n}} &
\colhead{$\dot{M}$\tablenotemark{o}} &
\colhead{$r_{\rm scat}$\tablenotemark{p}} 
\\
\colhead{\#} &
\colhead{$(M_J)$} &
\colhead{$(R_J)$} & 
\colhead{$(\rm AU)$} &
\colhead{$(R_J)$} & 
\colhead{$(\rm \mu bar)$} &
\colhead{$(\rm km/s)$} & 
\colhead{$(\rm G)$} &
\colhead{} &
\colhead{} &
\colhead{} &
\colhead{($R$)} &
\colhead{($R$)} &
\colhead{($R$)} &
\colhead{} &
\colhead{$\rm (g/s)$} &
\colhead{$(R)$}
}
\startdata
1& 0.7 & 1.3 & 0.05 & 1.4 & 0.04 & 10.0 & 8.6 & 9.3 & 0.032 & 0.054 & 4.6  & 3.1 & 6.0 & 0.074 & $ 9.3 \times 10^{10}$ & 6.6 \\
2& 0.7 & 1.3 & 0.05 & 1.4 & 0.4 & 10.0 & 8.6 & 9.3 & 0.032 & 0.54 & 4.6  & 3.1 & 3.3 & 0.31 & $ 1.4 \times 10^{12}$ & 10.0 \\
3& 0.7 & 1.3 & 0.05 & 1.4 & 0.004 & 10.0 & 8.6 & 9.3 & 0.032 & 0.0054 & 4.6 & 3.1 & 9.7 & 0.025 & $6.6 \times 10^9$ & 4.5 \\
4& 0.7 & 1.3 & 0.05 & 1.4 & 0.04 & 13.0 & 8.6 & 5.5 & 0.019 & 0.054 & 4.6 & 1.8 & 3.0 & 0.24 & $1.7\times 10^{12}$ & 9.7 \\
5& 0.7 & 1.3 & 0.05 & 1.4 & 0.04 & 7.5 & 8.6 & 16.6 & 0.056 & 0.054 & 4.6 & 5.5 & 23.2 & 0.030 & $1.5\times 10^8$ & 2.8 \\
6& 0.7 & 1.3 & 0.05 & 1.4 & 0.04 & 10.0 & 43.0 & 9.3 & 0.032 & 0.0022 & 4.6 & 3.1 & 11.6 & 0.091 & $5.9\times 10^{10}$ & 6.8 \\
7& 0.7 & 1.3 & 0.05 & 1.4 & 0.04 & 10.0 & 2.9 & 9.3 & 0.032 & 0.49 & 4.6 & 3.1 & 3.4 & 0.072 & $1.4\times 10^{11}$ & 7.8 \\
8& 0.7 & 1.3 & 0.025 & 1.4 & 0.04 & 10.0 & 8.6 & 9.3 & 0.25 & 0.054 & 2.3 & 3.1 & 6.4 & 0.075 & $6.6\times 10^{10}$ & 4.2 \\
9& 0.7 & 1.3 & 0.10 & 1.4 & 0.04 & 10.0 & 8.6 & 9.3 & 0.0040 & 0.054 & 9.2 & 3.1 & 5.9 & 0.096 & $1.4 \times 10^{11}$ & 7.7
\enddata
\tablenotetext{a}{ Planet mass. }
\tablenotetext{b}{ Transit radius. }
\tablenotetext{c}{ Orbital separation. }
\tablenotetext{d}{ Radius at the base of the warm H layer.}
\tablenotetext{e}{ Pressure at base radius ($r=R$) at the substellar point
$(\theta,\phi)=(\pi/2,0)$. }
\tablenotetext{f}{ Isothermal sound speed.}
\tablenotetext{g}{ Magnetic field at pole ($\theta=0,\pi$) at base radius ($r=R$).}
\tablenotetext{h}{ $\lambda=GM_p/Ra^2$ }
\tablenotetext{i}{ $\epsilon=(\Omega R/a)^2$, where $\Omega$ is the orbital frequency.}
\tablenotetext{j}{ $\beta=8\pi P_{\rm ss}/(B_0/2)^2$ is the base value at the substellar point. }
\tablenotetext{k}{ Radius of $L_1-L_2$ Lagrange points. }
\tablenotetext{l}{ Sonic point radius ignoring tides and field line curvature (eq.\ref{eq:rs0}).}
\tablenotetext{m}{ Dead zone radius in $\phi=\pi/2$ plane. }
\tablenotetext{n}{ Integrated Lyman $\alpha$ transit depth from $-200$ to $200\ {\rm km\ s^{-1}}$
from line center. The disk inside the base radius is asssumed opaque, and contributes $(R/R_\star)^2
\simeq 0.015$ to the transit depth for $R=1.4 R_J$ and $R_\star=1.15 R_\odot$. }
\tablenotetext{o}{ Computed mass loss rate.}
\tablenotetext{p}{ Radius $(A_{\rm scat}/\pi)^{1/2}$ corresponding to area $A_{\rm scat}$ in $x-z$ plane over which $\tau_\nu \geq 1$
for Lyman $\alpha$ cross section $\sigma_\nu=10^{-15}\ {\rm cm^2}$. This cross section corresponds to a velocity $ \simeq 25\ {\rm km\
s^{-1}}$ from line center.}
\end{deluxetable*}

Detailed results will be presented for the 9 models listed in Table
\ref{tab:models}. The planetary mass and radius, and the stellar mass
and radius are characteristic of HD 209458b. We vary parameters not
directly measured, such as $P_{\rm ss}$, $a$, $B_0$, as well as the
orbital radius $D$.

The numerical implementation of the sonic point solver deserves
further discussion.  Anywhere from zero to several solutions to
eq.\ref{eq:nozzleeqn} may be found in the interval $R \leq r \leq
r_{\rm max}$. Some sonic points may be spurious if a potential barrier
exterior to the sonic point decelerates the flow to subsonic, even
zero, speed. These spurious solutions are discarded by defining the
true sonic point solution to be a {\it global} minimum of $-\ln(B/B_0)
- U/a^2$ which occurs {\it within} the interval $R < r < r_{\rm max}$.
If the global minimum occurs at either of the endpoints of the interval,
then there is no good sonic point solution.

For minimum at the base $r=R$, the integration is flagged as a region of
parameter space with no solution, as we should have used a base position
deeper in the planet. For sonic points sufficiently deep in the planet that
ram pressure dominates magnetic pressure at the sonic point, we expect
the Roche lobe overflow model to be recovered. The other problem is that
the global minimum can occur at the outer boundary of the integration,
$r=r_{\rm max}$. For instance, this can occur in the polar regions due
to the upwardly increasing potential. For $r>r_d$, field lines with
$\sin\theta>\sin\theta_{\rm crit}= f^{-1/2}$ have outward tidal force.
Field lines with outward tidal force will have fluid accelerated outward,
promoting the existence of a sonic point.  The field line starting
at base co-latitude $\sin\theta_{b,crit}=(R/fr_d)^{1/2}$ will be the
last field line on which the tidal force points outward. Accordingly,
if no sonic point solution is found in the radial range $R < r < r_{\rm
max}$, and the field line has $\sin\theta<\sin\theta_{b,crit}$, then we
treat the field line as hydrostatic and set the velocity to zero. Such
field lines have outwardly increasing potential, and hence outwardly
decreasing density.  If, however, we had used a field line model that
allowed the field at $r>r_d$ to bend downward toward the orbital plane,
it is likely that a sonic point could have been found. This affects our
later numerical results, as we analytically predicted the critical tidal
strength to be $\epsilon = 4/\lambda^2$ in order to find sonic points in
the polar region, whereas the above prescription would force these field
lines to be hydrostatic due to the potential barrier.  This approximate
treatment of the polar regions likely does not affect either the total
mass loss rate, or the column density profiles, since the sonic point
will be so far from the planet that the velocity near the planet is
quite subsonic, and the fluid will be nearly hydrostatic.

\begin{figure}[t]
\epsscale{1.2}
%\plotone{fieldline_rhov_vs_r.ps}
\plotone{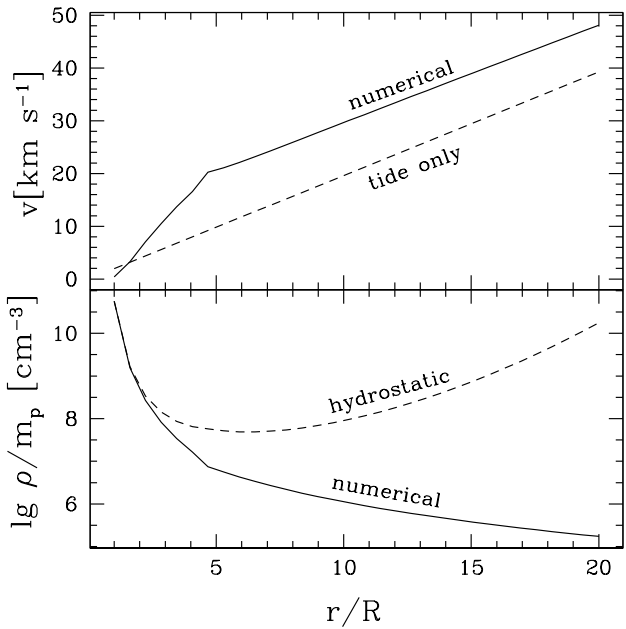}
\caption{ Velocity and density versus radius along a field line in the wind
zone. Model 1 parameters from Table \ref{tab:models} are used. The field line is located at
$\phi=0$ with $\theta_b=0.35\ {\rm rad}$. The dead zone is at $\theta_d=0.48\ {\rm rad}$
and the critical field line is at $\theta_{\rm b, crit}=0.23$. The discontinuity
at $r/R=4.7$ is the dead zone radius, where the field changes shape. The line labeled
``tide only" is the asymptotic approximation in eq.\ref{eq:eqforv},
and the line labeled ``hydrostatic" evaluates the density using the hydrostatic
balance approximation in eq.\ref{eq:isoP}.
}
\label{fig:fieldline}
\end{figure}

Figure \ref{fig:fieldline} shows the density and speed along a
particular field line in the wind zone. The change in slope near $r=4.7R$
is the change in field geometry at $r=r_d$. The speed along the field
line is somewhat larger than the asymptotic result in eq.\ref{eq:vasymp}
due to the enthalpy ($\ln$) terms in eq.\ref{eq:eqforv}.  In the lower
panel, the density is approximately hydrostatic inside the sonic
point at $r_s=2.4R$, and decreases roughly as $\rho \propto B/v \propto
r^{-3}$ outside that point. This plot explicitly demonstrates that the
gas density in the wind zone can be orders of magnitude smaller than
nearby gas in the dead zone, which satisfies hydrostatic balance.

%\begin{figure}[t]
%\epsscale{1.2}
%\plotone{density_x=0_case1.ps}
%\plotone{density_y=0_case1.ps}
%%\plottwo{density_x=0_case1.ps}{density_y=0_case1.ps}
%\caption{ Contours of total number density, $\log_{10}\rho/m_p[\rm cm^{-3}]$
%in the y-z plane (upper) at $x=0$, and in the x-z plane (lower) for $y=0$.
%Model 1 parameters are used (see Table \ref{tab:models}).
%}
%\label{fig:density}
%\end{figure}

\begin{figure}[t]
\epsscale{1.2}
\plotone{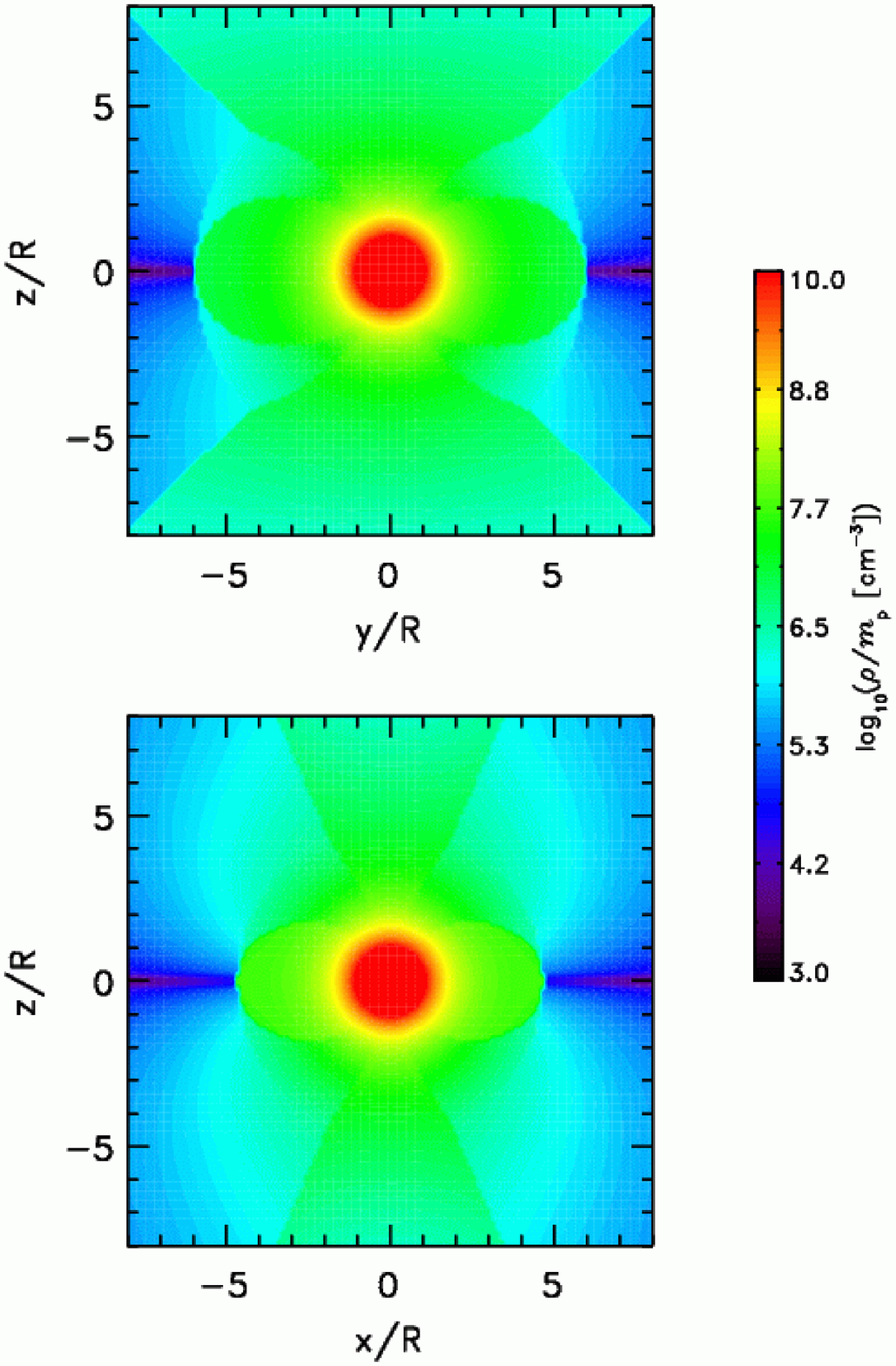}
\caption{ Contours of mass density, $\log_{10}\rho/m_p[\rm cm^{-3}]$
in the $y-z$ plane (upper) at $x=0$, and in the $x-z$ plane (lower) for $y=0$.
Model 1 parameters are used (see Table \ref{tab:models}).
}
\label{fig:density}
\end{figure}

Figure \ref{fig:density} shows contours of mass density on slices through
the center of the planet in the $y-z$ plane, as viewed during transit,
and the $x-z$ plane, as viewed midway between primary and secondary
transit. The quantity plotted is $\rho/m_p = n_{\rm H} + n_p$ (note
that this quantity is distinct from the total number density $n_{\rm
tot} = n_{\rm H} + 2 n_p$, which depends on the
details of the photoionization model). Model 1 parameters listed in
Table \ref{tab:models} were used.  Near the planet the contours are
approximately spherical, since the velocities are everywhere subsonic
and the tidal force is small.  The bulge at the equator is the equatorial
dead zone.  The poles are hydrostatic as tides have shut down the wind,
and the inward tidal force at the pole causes the density to decrease
outward faster than at the equator.  The impact of the tidal force on
the dead zone can be seen by comparing the upper and lower panels in
Figure \ref{fig:density}.  Along the $x$-direction, the outward tidal
force decreases the size of the dead zone, but the same tidal force also
causes the dead zone to have higher density.  Outside the sonic point,
the density in the wind zone is smaller than in the neighboring dead zone,
which pushes the density contours inward in the wind zone.  Near the
equator, outside the dead zone, the density becomes quite small, hence
the pile-up of contours near the critical angle $\sin\theta_{\rm crit}$.

In the next section, the wind models in Table 1 are used to compute the
planetary mass loss rate in the wind zone.

%%%%%%%%%%%%%%%%%%%%%%%%%%%%%%%%%%%%%%%%%%%%%%%%%%%%%%%%%%%%%%%%%%%%%

\section{ mass loss rate and spin-down torque}
\label{sec:mdot}

\begin{figure}[t]
\epsscale{1.2}
\plotone{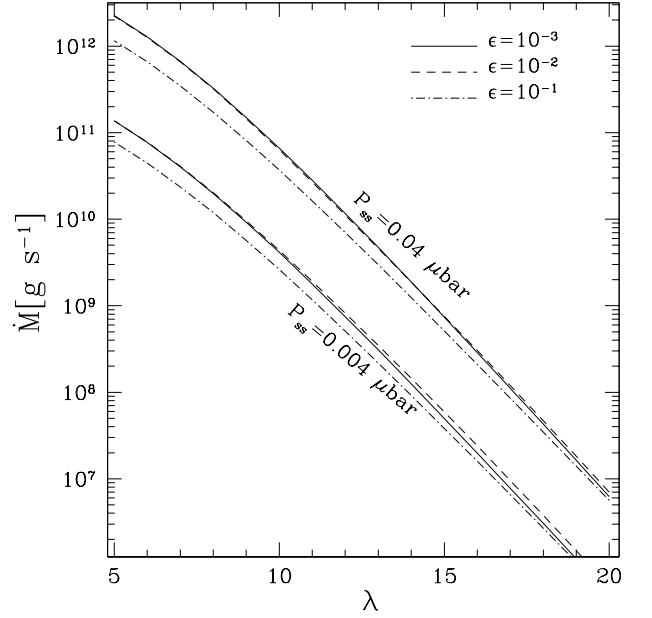}
\caption{ Total mass loss rate as a function of $\lambda$ for $M_p=0.7
M_J$, $R=1.4 R_J$ and $B_0=8.6\ {\rm G}$. The upper (lower) set of three curves
uses base pressure $P_{\rm ss}=0.04\ (0.004) \mu {\rm bar}$.
The line style in each set of three curves gives the value of $\epsilon$, 
the tidal strength.
}
\label{fig:mdot}
\end{figure}

The mass loss rate is computed by integrating $\rho v_r=\rho b_r v_b$ over
the surface area of the wind zone at the base. Using the base density 
from eq.\ref{eq:rhob},
the mass loss rate is
\be
\dot{M} & = &  R^2 \rho_{ss} \int_0^{2\pi} d\phi \int_{\rm wind\ zone} d\theta_b \sin\theta_b\  b_r v_b
\left( \frac{ \rho_b(\theta_b,\phi)}{\rho_{ss}} \right)
\nonumber \\ & = &  R^2 P_{ss} a^{-1}\  {\cal F}(\lambda,\epsilon,\beta)
\nonumber \\ & = & 
4.0 \times 10^{12}\ {\rm g\ s^{-1}}\
\left( \frac{R}{1.4R_J} \right)^2
\nonumber \\ & \times & \left( \frac{P_{ss}}{\rm 0.04 \mu bar} \right) \left( \frac{10\ \rm km\ s^{-1}}{a}
\right)\ {\cal F}(\lambda,\epsilon,\beta),
\ee
where the dimensionless integral
\be
{\cal F}(\lambda,\epsilon,\beta) & = & 8 \int_0^{\pi/2} d\phi \int_0^{\theta_d} d\theta_b \sin\theta_b\
b_r \left( \frac{v_b}{a} \right)
\left( \frac{ \rho_b(\theta_b,\phi)}{\rho_{ss}} \right).
\ee
The mass loss rate for fixed $M$, $R$ and $B_0$, but varying $a$, $D$
and $P_{ss}$ is shown in Figure \ref{fig:mdot}.  The steep decline of $\dot{M}$
with $\lambda$ is due to smaller density at the sonic point radius. The
mass loss decreases slightly for large $\epsilon$ due to the smaller
fraction of open field lines. 

Why is the mass loss rate $\dot{M}$ proportional to the base
pressure $P_{ss}$? Recall that we are approximating the true atmosphere
with an isothermal model. The appropriate values of $a^2$ and $P_{ss}$,
as determined by photoionization equilibrium and heating/cooling balance,
have been discussed in section \ref{sec:Hlayer}.  The sonic point lies at
a fixed radius, given roughly by eq.\ref{eq:rs0} based on the choice of
sound speed $a$.  The location of the base of the isothermal layer is also
at a fixed radius, estimated to be 1.1$R_{\rm ph}$ (see \S~\ref{sec:pi}).
By eq.\ref{eq:isoP}, $\rho (r=r_s) \propto \rho_{ss}$, therefore the
density at the sonic point is proportional to the base density $\rho_{ss}$
(and therefore also $P_{ss}$).  Consequently, if a larger value of $P_{ss}$
is required to explain the transit depth, the mass loss must be
increased proportionally.

The mass loss rates in Figure \ref{fig:mdot} are largely consistent with
previous studies \citep[e.g.,][]{2009ApJ...693...23M} when comparable gas
density is used. By comparison, an unmagnetized, spherically symmetric,
isothermal wind would have \citep{1999isw..book.....L} ${\cal F} \simeq
\pi \lambda^2 \exp(3/2-\lambda)$, which would be a factor of $\simeq 3-10$
larger than the curves in Figure \ref{fig:mdot}, and with a slightly
flatter slope. Inclusion of the magnetic field decreases the mass loss
rate, mainly due to the decrease in area occupied by the wind zone.

The angular momentum loss rate depends on the radius at which the
torque is applied. For an isolated
planet, the field lines remain rigid out to the Alfv\'en radius. But this location may be at many tens of planetary
radii, and may be pre-empted by the interaction of the planetary wind
with the stellar wind. By assuming the torque is exerted at a radius
$r_{\rm torque}$ we estimate an angular momentum loss rate
\be
\dot{M} \Omega r_{\rm torque}^2
& \simeq &
7.3 \times 10^{28}\ {\rm erg}\ \left( \frac{1\ \rm day}{P_{\rm orb}} \right)
\nonumber \\ & \times & 
\left( \frac{ r_{\rm torque}} {10^{11}\ {\rm cm}} \right)^2 \left( \frac{\dot{M}}{10^{11}\ {\rm g\ s^{-1}} }
\right).
\label{eq:Jdot}
\ee
While this torque may cause moderate changes in the spin rate for an
isolated planet on Gyr timescales, it likely not large enough to torque
the planet away from synchronous rotation to the extent that significant
gravitational tidal heating will occur \citep[see][for a discussion of the necessary torques]{2009arXiv0901.0735A}.

%%%%%%%%%%%%%%%%%%%%%%%%%%%%%%%%%%%%%%%%%%%%%%%%%%%%%%%%%%%%%%%%%%%%%

\section{ neutral hydrogen column density}
\label{sec:NH}

Given the global MHD models derived in section \ref{sec:global},
we require a model for the run of ionization in the magnetosphere in
order to compute observable quantities such as the transmission spectrum.
In section \ref{sec:Hlayer}, we discussed a model including only the
dominant processes: photoionization and radiative recombination of
hydrogen. In the remainder of the paper, we use the simpler optically thin
limit to evaluate $n_H$ given $\rho$:
\be
n_H & = & \left( \frac{\sqrt{ n_{\rm eq,0} + 4 \frac{\rho}{m_p} } - \sqrt{n_{\rm eq,0}} }{2} \right)^2.
\label{eq:nH}
\ee
Here $n_{\rm eq,0}=J_0/\alpha_B$ is the density at which $n_H=n_p$ in
the optically thin limit. The use of $J_0$ instead of $J$ simplifies
the calculation, as only the local density is required, and not the
column $N_H$. This approximation may underestimate $n_H$ near the
H-H$^+$ transition, but should be adequate for our purposes.

%\begin{figure*}[t]
%\epsscale{0.38}
%\plotone{column_case1.ps}
%\plotone{column_case2.ps}
%\plotone{column_case3.ps}
%\plotone{column_case4.ps}
%\plotone{column_case5.ps}
%\plotone{column_case6.ps}
%\plotone{column_case7.ps}
%\plotone{column_case8.ps}
%\plotone{column_case9.ps}
%\caption{
%Contours of hydrogen column density, $\log_{10}(N_H[cm^{-2}])$,
%versus impact parameter in the $y-z$ plane observed during transit.
%The parameters for each plot are given in table \ref{tab:models}.
%The top row is cases 1-3, from left to right. The middle row is cases 4-6,
%from left to right.  The bottom row is cases 7-9, from left to right.
%}
%\label{fig:NHcontours}
%\end{figure*}

\begin{figure*}[t]
\epsscale{1.0}
\plotone{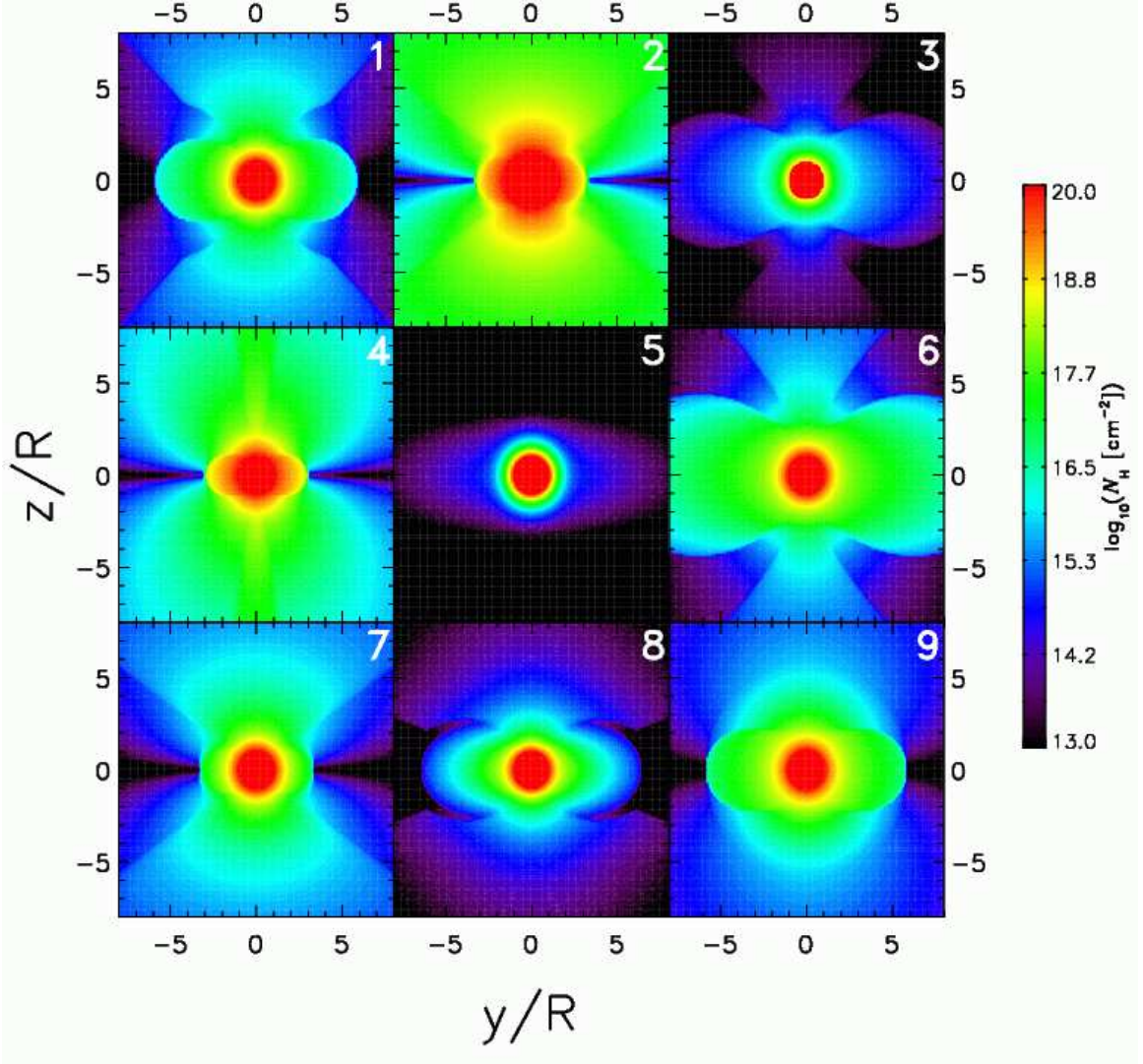}
\caption{
Contours of hydrogen column density, $\log_{10}(N_H[{\rm cm}^{-2}])$,
versus impact parameter in the $y-z$ plane observed during transit.
The parameters for each plot are matched to the labeled model numbers
in Table \ref{tab:models}.
}
\label{fig:NHcontours}
\end{figure*}

To evaluate the neutral hydrogen column density, we first evaluate $\rho$
(\S~\ref{sec:global}), and then $n_H$ (eq.\ref{eq:nH}), on a grid of
$(x,y,z)$, with each coordinate in the range $(-1.1R_\star,1.1R_\star)$.
The column is displayed as seen at transit, i.e.~ we integrate over the
coordinate along the star planet line to get column
\be
N_H(y,z) & = & \int_{-1.1R_\star}^{1.1R_\star} dx\ n_H(x,y,z)
\ee
as a function of the impact parameters $y$ and $z$.  Figure
\ref{fig:NHcontours} shows contours of hydrogen column density for the
9 models listed in Table \ref{tab:models}.  All models have $M_p=0.7M_J$
and $R=1.4R_J$, and vary a single parameter $P_{\rm ss}$, $a$, $B_0$ and
$D$ in turn.  The fiducial case, Model 1, clearly shows the equatorial and
polar dead zones, as well as the mid-latitude wind zone with comparatively
smaller $N_H$.  Model 2 (3) has $P_{ss}$ larger (smaller) by a factor of
10. This has the effect of decreasing (increasing) the dead zone size
as well as scaling up (down) the density in the dead zone.  Model 4 (5)
has larger (smaller) $a$, leading to larger (smaller) density at a given
distance from the planet, as well as increasing (decreasing) the size of
the dead zone. Model 6 (7) has larger (smaller) $B_0$, which increases
(decreases) the size of the dead zone. Model 8 (9) has smaller (larger)
$D$. Larger tide is more effective in shutting down the wind at
the pole, but also decreases the size of the dead zone.

Aside from the overall magnitude mainly set by the base pressure $P_{ss}$
and sound speed $a$, the dominant parameters determining the appearance
of each plot are the equatorial and polar dead zone sizes. The equatorial
dead zone size (see Figure \ref{fig:cusp}) depends on $\lambda$, $\beta$
and $\epsilon$.  The size of the polar dead zone is set by the strength
of the tidal force. The critical tidal strength in eq.\ref{eq:epscrit}
refers to shutting down the wind at $\theta_b=0$, and assumes dipole field
geometry. For $\epsilon > \epsilon_{\rm crit}$, a range of $\theta_b$
near the pole can have the wind shut off. The result depends on the which
field geometry is chosen. For instance, the sonic point eq.\ref{eq:speqn}
can be rederived for radial field lines. The discriminant of this cubic
equation can be used to show that no sonic point can be found for
\be
\sin^2 \theta & \la & \frac{1}{f\epsilon} \left( \epsilon - \frac{32}{27\lambda^2}
\right).
\ee
The critical tidal strength for radial field lines is $\epsilon_{\rm
crit}=32/(27\lambda^2)$, a slightly different numerical coefficient than
the dipole case. As $\epsilon$ increases above $\epsilon_{\rm crit}$,
the size of the polar dead zone increases. In the limit $\epsilon \gg \epsilon_{\rm
crit}$, the wind is shut down in the entire region $\sin^2\theta \leq 1/f$ where
the tidal force is inward. For large $\epsilon$ and small $\beta$, 
the polar and equatorial dead zones can dominate the volume near the planet (e.g., Model 6). 

%%%%%%%%%%%%%%%%%%%%%%%%%%%%%%%%%%%%%%%%%%%%%%%%%%%%%%%%%%%%%%%%%%%%%

\section{ Lyman $\alpha$ transmission spectra }
\label{sec:lymanalpha}

In section \ref{sec:NH} we focused on understanding the hydrogen column
as a function of impact parameter, including the dependence on unknown
parameters such as temperature and magnetic field. An additional effect on the transmission spectrum is the velocity
gradients in the wind, which were studied in sections \ref{sec:wind}
and \ref{sec:global}. In this section we compute the Lyman $\alpha$
transmission spectra for the global models, including both column and
velocity gradient effects.

The transmission function, $T_\nu$, is the fraction of stellar flux at
frequency $\nu$ which passes through the planet's atmosphere without
suffering scattering out of the beam. In terms of the out-of-transit
stellar flux, $F_\nu^{(0)}$, and the in-transit flux, $F_\nu$,
$T_\nu$ is defined as
\be
T_\nu & = & \frac{ F_\nu }{ F_\nu^{(0)} }.
\label{eq:Tnuobs}
\ee
If the interstellar medium (ISM) optical depth, $\tau_\nu^{\rm (ISM)}$,
is constant over the stellar disk, and in time, and the geocoronal
emission is independent of time, then the ratio in eq.\ref{eq:Tnuobs}
depends solely on the properties of the planetary atmosphere, and is the
fundamental quantity to compare to the data.

We compute $T_\nu$ as follows.  Let $\sigma_\nu$ be the Lyman $\alpha$
line cross section.  We simplify the problem by assuming the planet to
be at the center of the stellar disk.  The optical depth through the
planet's atmosphere at position $(y,z)$ on the stellar disk is
\be
\tau^{\rm (p)}_\nu(y,z) & = & \int dx\ n_H(x,y,z)\ \sigma_\nu(x,y,z).
\ee
Assuming the stellar intensity is uniform over the disk,
\be
T_\nu & = & \frac{1}{\pi R_\star^2} \int dy dz\ e^{- \tau^{\rm (p)}_\nu(y,z) },
\label{eq:Tnu}
\ee
where the integral extends over $y^2+z^2 \leq R_\star^2$.
As an integrated measure of the transit depth, we compute
\be
\frac{\delta F}{F} & = & \frac{ \int d\nu I_\nu^{(\star)} e^{- \tau_\nu^{\rm (ISM) } }
\left( 1 - T_\nu \right) }
{ \int d\nu I_\nu^{(\star)} e^{- \tau_\nu^{\rm (ISM) } } }
\label{eq:dFoverF}
\ee
where $I_\nu^{(\star)}$ and $\tau_\nu^{\rm (ISM) }$ are the unabsorbed
stellar intensity and ISM optical depth, both assumed uniform over the
disk.  In practice, we follow \citet{2008ApJ...688.1352B} and integrate
over $- 200\ \rm km\ s^{-1} \leq \Delta v \leq 200 \rm km\ s^{-1}$.
The interstellar medium (ISM) is assumed to have a temperature $T_{\rm
ism}=8000\ {\rm K}$ and hydrogen column $N_{\rm H, ism} = 10^{18.4}\
{\rm cm^{-2}}$ \citep{2005ApJS..159..118W}, implying the line is dark
inside linewidth $\Delta v = c(\nu-\nu_0)/\nu_0 \la 50\ \rm km\ s^{-1}$.
For $I_\nu^{(\star)}$ we use the following (unnormalized) fit to the quiet
solar Lyman $\alpha$ spectrum presented in \citet{1997ApJS..113..195F}
(downloaded from http://www.mps.mpg.de/projects/soho/sumer/FILE/
Atlas.html):
\be
I_\nu^{(*)} & = &
 \left[ 1 + \left( \frac{\Delta v}{67\ {\rm km\ s^{-1}}} \right)^3 \right]^{-1}.
\label{eq:sumer}
\ee 
The Voigt function $H(a,u)$ \citep{1978stat.book.....M} is used for the line profile, giving
\be
\sigma_\nu & = & \frac{\pi e^2}{m_e c} f_{12} \frac{1}{\sqrt{\pi} \Delta \nu_D}
H(a_D,u)
\ee
where $-e$ is the electron charge, $m_e$ is the electron mass, and
$f_{12}=0.42$, $\lambda_0=1215$\AA \ and $\nu_0=c/\lambda_0$
are the Lyman $\alpha$ oscillator strength, line center wavelength and
frequency. The Doppler width is $\Delta \nu_D=\nu_0 v_{\rm th}/c$
where the hydrogen atom thermal velocity is $v_{\rm th}=(2k_bT/m_p)^{1/2}$.
The damping parameter is $a_D=\Gamma/4\pi \Delta \nu_D$, where 
the natural linewidth is $\Gamma=6.25 \times 10^8\ {\rm s^{-1}}$.
Finally, the distance from line center, in Doppler widths,
including both bulk motion and thermal broadening, is
\be
u & = & \frac{\nu-\nu_0}{\Delta \nu_D} + \frac{v_x}{v_{\rm th}}
\ee
where $v_x$ is the bulk motion directed from planet toward star, which is
away from the observer.

%\begin{figure}[t]
%\epsscale{1.0}
%\plotone{xsec_lymanalpha.ps}
%\caption{ Lyman $\alpha$ cross section as a function of velocity width from
%line center. In each panel, the three lines correspond to
%temperature $T[K]=10^4$ (solid), $10^5$ (dotted) and $10^6$ (dashed).
%The three panels are for bulk velocity away from
%the observer $v_x[km\ s^{-1}]=0$ (upper),
%50 (middle) and 100 (lower). For bulk velocity toward the observer, the curves
%would be reflected about zero to positive velocities.
%}
%\label{fig:lymanalphaxsec}
%\end{figure}
\begin{figure}[t]
%\epsscale{1.0}
\epsscale{1.1}
\plotone{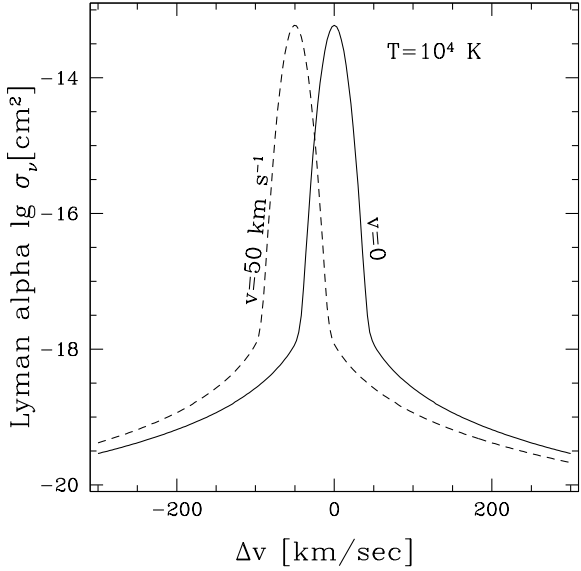}
\caption{ Lyman $\alpha$ cross section as a function of velocity
width from line center. Both profiles are given for a temperature $T=10^4$
K.  The solid line is for zero bulk velocity away from the observer, while
the dashed line is for $\Delta v=50$ km s$^{-1}$. For bulk velocity toward
the observer, the dashed curve would be reflected about $\Delta v=0$.
}
\label{fig:lymanalphaxsec}
\end{figure}

There are three instructive limits of eq.\ref{eq:Tnu} to guide the intuition.
First, if $v_x=0$ and the gas is optically thick over an area $A_{\rm tran}$, with
negligible optical thickness outside this area, the fraction of flux absorbed
by the planet is
\be
1 - T_\nu & =& \frac{A_{\rm tran}}{\pi R_\star^2}
= 0.013\ \left( \frac{ A_{\rm tran} }{\pi (1.3R_J)^2} \right)
\left( \frac{1.15R_\odot}{R_\star} \right)^2.
\ee
Next, if $v_x=0$ and the gas is optically thin, then
the transit signal due to the optically thin area is
\be
1 - T_\nu & \simeq &
\frac{1}{\pi R_\star^2} \int_{\tau^{\rm (p)}_\nu \ll 1} dy dz \tau_\nu^{\rm (p)}(y,z)
\equiv \langle \tau_\nu^{\rm (p)} \rangle,
\ee
which is just the area-averaged optical depth,
and is proportional to the total number of hydrogen atoms
times their mean cross section. The third limit is when thermal motions
are much smaller than bulk motions, and the line profile can be approximated
as a delta function $\delta \left[ \nu - \nu_0(1-v_x/c) \right]$. In this case,
the cross section is only nonzero at those values of $x_\star=x_\star(\nu,y,z)$ where 
$v_x(x_*,y,z)=c(\nu_0-\nu)/\nu_0$ is satisfied, so that the photon is shifted to
line center in the atom's frame. In this case the optical depth becomes
\be
&& \tau_\nu^{\rm (p)}(y,z)  =  n_H(x_\star,y,z) \frac{\pi e^2}{m_e c} f_{12} \frac{\lambda_0}{|\partial
v_x(x_\star,y,z)/\partial x|}
\nonumber \\ & \simeq  & 
 2 \times 10^{-3} \left( \frac{n_H(x_\star,y,z)}{\rm 1\ cm^{-3}} \right)
\left( \frac{ P_{\rm orb} }{\rm 1\ day} \right)
\frac{\Omega}{|\partial v_x(x_\star,y,z)/\partial x|},
\label{eq:taubulk}
\ee
where in the second equality we have scaled the velocity gradient to
the orbital frequency $\Omega$. Eq.\ref{eq:taubulk} shows
that for hydrogen densities $n_H \ga 10^{2-3}\ {\rm cm^{-3}}$,
the optical depth along a line of sight will be high provided that
there is gas with sufficiently large velocity to absorb at that wavelength.

Figure \ref{fig:lymanalphaxsec} shows the cross section as a function
of frequency in velocity units, at $T=10^4$ K and for $\Delta v=
0$ and $\Delta v=50 $ km s$^{-1}$. For HD 209458b, the transit radius is $R_{\rm ph}=1.3R_J$
and the stellar radius is $R_\star=1.15R_\odot$, giving a transit
depth $\delta F/F = 0.013$ in the optical continuum.  To explain the
line-integrated Lyman $\alpha$ transit depth $ \simeq 9\%$ \citep[e.g., see the discussion in][]{2008ApJ...688.1352B} one could invoke an
opaque disk of area $\sim \pi (2.6R_{\rm ph})^2$.  The central issue is
that this disk must be opaque at $\Delta v \ga \pm 100\ {\rm km\ s^{-1}}$
from line center, requiring large columns of neutral hydrogen at radii
$2-3R_{\rm ph}$.

In Figure \ref{fig:pieq}, triangle symbols show where
Lyman $\alpha$ radiation at frequencies $\pm 100\ {\rm km\ s^{-1}}$ from
line center is optically thick on a radial line outward. This point is
much deeper in the atmosphere from where Lyman continuum at threshold
becomes optically thick, due to the rapid decrease in Lyman $\alpha$
cross section.  Clearly in order to model the transit
spectrum in the wavelength region of interest, one must include regions
down to $\sim 1-10\ \rm nbar$ in the atmosphere. To quantify this statement, we compute the optical depth through the H layer
where $n_H \simeq \rho/m_p$ is given by eq.\ref{eq:isoP0}. Assuming the dominant
contribution arises from the layer of steeply falling density, the slant optical
depth is dominated by the region near $x=0$ and we find
\be
&& \tau_\nu^{\rm (p)}(y,z) = \sigma_\nu \left( \frac{\rho(x=0,y,z)}{m_p} \right)
\nonumber \\ & \times &
\int dx 
\exp \left[ - \frac{1}{2a^2} \left( \frac{GM_p}{b^3} - 3\Omega^2 \right) x^2 \right]
\nonumber \\ & \simeq & \sigma_\nu 
\left( \frac{\rho(x=0,y,z)}{m_p} \right) \left( \frac{2\pi b^3/\lambda R}{  1 - (b/r_L)^3 }
\right)^{1/2}
\nonumber \\ & \simeq & 
1.2 \left( \frac{ 100\ \rm km\ s^{-1} }{\Delta v} \right)^2 
\left( \frac{P(x=0,y,z)}{1\ \rm nbar} \right)
\left( \frac{ 10\ \rm km\ s^{-1} }{a} \right)^2
\nonumber \\ & \times & 
\left( \frac{10}{\lambda} \right)^{1/2}
\left( \frac{b}{R} \right)^{3/2} \left( \frac{R}{1.3\ R_J} \right)
\left( 1 - (b/r_L)^3 \right)^{-1/2}.
\label{eq:tauan}
\ee
Here $b=\sqrt{y^2+z^2}$ is the impact parameter, eq.\ref{eq:U} was used
for the tidal potential, $r_L$ is given by eq.\ref{eq:rL}, and the last
equality assumes the cross section is on the damping wing (see Figure
\ref{fig:lymanalphaxsec}). In the H$^+$ layer, eq.\ref{eq:tauan} should
be multiplied by $1/2$ to account for the smaller H atom scale height.
Eq.\ref{eq:tauan} agrees roughly with the position of the triangles in
Figure \ref{fig:pieq}, keeping in mind that the slant length is a factor
of a few larger than the scale height. Eq.\ref{eq:tauan} shows that 
the Lyman $\alpha$ transmission spectrum at $\Delta v=\pm 100\ \rm km\ s^{-1}$
is probing down to $\la \rm nbar$ pressures, depending on the
value of $b/R$. 

\begin{figure*}[t]
\plotone{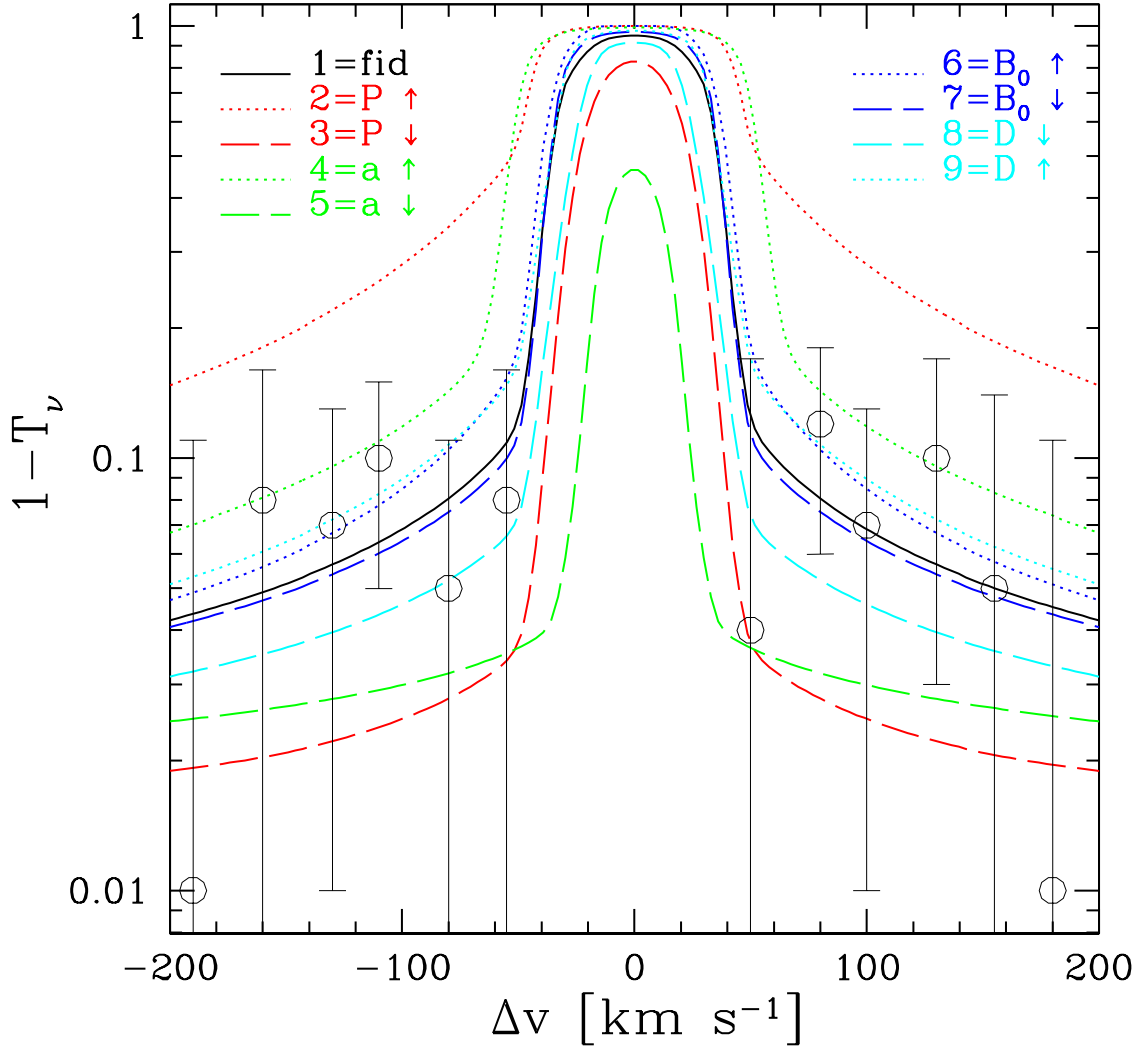}
\caption{
Fractional flux decrease, $1-T_\nu$, versus frequency in velocity units
for HD 209458b.  Curves for Models 1-9 from Table \ref{tab:models} are
computed from eq.\ref{eq:Tnu} and points with error bars are the data
from Figure 6 of \citet{2008ApJ...688.1352B}.
}
\label{fig:absspec}
\end{figure*}

To give a more precise numerical estimate of the transit depth, we first
compute the integrated quantity $\delta F/F$ as in eq.\ref{eq:dFoverF}
for the 9 models in Table \ref{tab:models}.  The result is given in
the table. The velocity range is taken to be $-200 \leq \Delta v[\rm
km\ s^{-1}] \leq 200$.  Since $1-T_\nu$ decreases away from line
center, $(1-T_\nu) F_\nu^{(0)}$ is peaked somewhat closer to line
center than $F_\nu^{(0)}$, the amount depending on the details of the
atmosphere. Transit depths of the correct magnitude $\delta F/F \sim
5-10\%$ can be achieved by adjusting the main parameters, $P_{\rm
ss} \simeq 10-100\ \rm nbar$ and $a \simeq 8-12\ \rm km\ s^{-1}$ to
have values as expected from the 1D model in Figure \ref{fig:pieq}.
The parameters $B_0$ and $D$ have a lesser impact by comparison.

The frequency dependent transit depth, $1-T_\nu$, was computed as in
eq.\ref{eq:Tnu} for the 9 models listed in Table \ref{tab:models},
and compared to the data for $(F_\nu^{(0)}-F_\nu)/F_\nu^{(0)}$ from
Figure 6 of \citet{2008ApJ...688.1352B}. The results are shown in
Figure \ref{fig:absspec}. Near line center, nearly the entire planetary
atmosphere is optically thick, and absorption is nearly complete. Moving
out from line center in the Doppler core, the cross section eventually
becomes small enough that part of the atmosphere becomes optically thin, after which
$1-T_\nu$ decreases rapidly. The curves level out when the damping wing
is reached, after which $1-T_\nu$ decreases slowly as the $\tau_\nu^{\rm
(p)} =1$ point moves deeper into the atmosphere as $P(x=0,y,z) \propto
\Delta v^2$. 

Given the large error bars, a range of parameter space agrees
with the data {\it if} the warm H layer extends sufficiently deep.
For instance, Model 3 with base pressure $P_{ss}=4\ \rm nbar$ is well
below the data points with the smallest error bars, in agreement
with \citet{2009ApJ...693...23M}.  The most sensitive parameter
dependencies are with the base pressure, $P_{\rm ss}$, and the sound
speed (temperature), $a$. Increasing the magnetic field has the effect of
increasing $1-T_\nu$ due to larger $N_H$ in the magnetosphere. Somewhat
offsetting this effect is that increasing $B_0$ decreases the size of the
wind zone, which decreases absorption near line center due to velocity
gradients. Perhaps counter-intuitively, moving the planet further from
the star increases the transit depth. Inspection of Figure \ref{fig:pieq}
shows that the H extends to both lower pressure and larger radius for more
distant planets with atmospheres in photoionization equilibrium. Lastly,
we note that velocity gradients are only important for $\Delta v \la 50\
\rm km\ s^{-1}$, and are more important for smaller $D$ due to the larger
tidal force.

We end this section with a brief discussion of scattering of Lyman
$\alpha$ from H atoms in the magnetosphere. The problem with observing
Lyman $\alpha$ during transit is that large $N_H$ is required to create
$\tau^{\rm (p)}_\nu \simeq 1$ at $\Delta v \ga 100\ {\rm km\ s^{-1}}$.
By contrast, at line center the cross section is $\sim 10^5$ times
larger, implying the atmosphere is optically thick at line center out to
much larger radii. We suggest that scattering of stellar Lyman $\alpha$
during the orbital phases in which the planet is moving toward or away
from the observer may be detectable, and provides a probe of thermal
gas in the magnetosphere, complementary to the transmission spectrum
measured during transit.  During the orbit, the Doppler shift of the
scattered spectrum varies in time due to the variation in line-of-sight
orbital motion. The orbital velocities naturally produces a feature in
the spectrum well outside the line core where ISM absorption dominates.

For a planet in circular orbit, there is no relative radial motion with
respect to the star, and the stellar spectrum at the planet is not Doppler
shifted. However, when an H atom in the planet resonantly scatters a
stellar photon, that H atom is moving with respect to the observer due
to the planet's orbital motion.  Photons emitted by the star near line
center ($\Delta v \la 67\ \rm km\ s^{-1}$) have their frequencies shifted
by $v_{\rm orb}=210\ {\rm km\ s^{-1}} (\rm 1\ day/P_{\rm orb})^{1/3}$
(for a solar mass star) due to the planet's orbital motion.  To assess the
area presented by the magnetosphere, we computed the area in the $x-z$
plane for which $\tau^{\rm (p)}_\nu \ga 1$ for $\sigma_\nu=10^{-15}\
\rm cm^2$, which corresponds to $\Delta v=25\
 \rm km\ s^{-1}$ from line center for $T=10^4\ \rm K$. The results are
tabulated in Table \ref{tab:models}. We find that the effective radius
of the scattering disk is $r_{\rm sc} \sim (5-10) R$ for the models
shown. The scattering disk for Lyman $\alpha$ is significantly larger than
the radius inferred during transit.
%A fraction 
%%$\pi r_{\rm sc}^2/2\pi
%%D^2 \simeq 0.028(r_{\rm sc}/10R_J)^2(1\ \rm day/P_{\rm orb})^{4/3}$ 
%$\pi r_{\rm sc}^2/4\pi
%D^2 \simeq 0.028(r_{\rm sc}/10R_J)^2(1\ \rm day/P_{\rm orb})^{4/3}$
%of line-center photons emitted by the star are scattered by the planet.
%Were this fraction to be interpreted as the scattering ``signal",
%relative to the stellar emission, it would be difficult to detect the
%scattered Lyman $\alpha$, even for the closest-in planets. However,
%inspection of eq.\ref{eq:sumer} shows that the line center flux is
%$\simeq 30(\Delta v/200\ \rm km\ s^{-1})^3$ times larger than that at
%$\Delta v$.  This increases the scattered flux signal relative to the
%background flux level.  Multiplying the fraction of photons scattered
%by the ratio of flux at line center to that at the orbital velocity,
%the expected change in flux, at $\Delta v=v_{\rm orb}$ from line center,
%is $ \delta F_\nu/F_\nu^{(0)} \simeq 0.8\ (r_{\rm scat}/10R_J)^2 (\rm 1\
%day/P_{\rm orb})^{7/3}$. While this signal may be small for HD 209458b
%at $P_{\rm orb}=3.5\ {\rm days}$ and $\delta F_\nu/F_\nu^{(0)}
%%\simeq 0.05$, 
%\simeq 0.03$,
%for planets with $P_{\rm orb}=1-2\ {\rm days}$ it may
%be large enough to be observable.
Assuming none of the resonantly scattered Lyman $\alpha$
photons are absorbed, and also assuming the Lambert phase function
\citep{1993tres.book.....H} as an estimate, the reflected flux is $F_{\rm
refl}(\nu) = F_\star(\nu^\prime) (2/3\pi) (r_{\rm sc}/D)^2$, where $\nu
\simeq \nu_0 \pm v_{\rm orb}$ is the observed frequency, $\nu^\prime
\simeq \nu_0$ was the frequency emitted by the star before Doppler
shift, and $F_\star(\nu)$ is the stellar Lyman $\alpha$ spectrum. The
size of the reflected flux relative to the flux emitted by the star out
on the wing at frequency $\nu$ is then $F_{\rm refl}(\nu)/F_\star(\nu) =
(F_\star(\nu^\prime)/F_\star(\nu)) (2/3\pi) (r_{\rm sc}/D)^2$. Inspection
of eq.\ref{eq:sumer} shows that the line center flux is $\simeq 30(\Delta
v/200\ \rm km\ s^{-1})^3$ times larger than that at $\Delta v$. This
acts to enhance the scattered flux signal relative to the background flux
level. Numerically we find the ratio of scattered, Doppler shifted flux
to background stellar flux is then $F_{\rm refl}(\nu)/F_\star(\nu) \simeq
0.4 (r_{\rm sc}/10R_J)^2 (1\ \rm day/P_{\rm orb})^{7/3}$.  While this
signal may be small for HD 209458b at $P_{\rm orb}=3.5\ {\rm days}$,
$F_{\rm refl}(\nu)/F_\star(\nu) \simeq 0.02 (r_{\rm sc}/10R_J)^2 $,
 for planets with $P_{\rm orb}=1-2\ {\rm days}$ it may
be large enough to be observable.

%%%%%%%%%%%%%%%%%%%%%%%%%%%%%%%%%%%%%%%%%%%%%%%%%%%%%%%%%%%%%%%%%%%%%

\section{ comparison to Roche-lobe overflow }
\label{sec:roche}

The magnetic wind model developed in this paper differs in
several respects from purely hydrodynamic mass loss models (e.g.,
\citealt{1975ApJ...198..383L}). In the standard Roche-lobe model
for nearly equal mass stars, nearly all the gas leaves the donor in a
narrow, cold stream through the L1 Lagrange point. The first assumption
underlying this solution is that $r_{s,0} \gg r_{\rm L1}$, so that the
gas is subsonic at the L1 equipotential for most ($\theta,\phi$). From
eq.\ref{eq:rL} and eq.\ref{eq:rs0}, this ratio is $r_{s,0}/r_L =
(\epsilon \lambda^2/9)^{1/3}$, and hydrodynamic Roche lobe overflow
requires $\epsilon \gg 9/\lambda^2$. Figure \ref{fig:eps_vs_lam} plots
$\epsilon$ versus $\lambda$, and shows that most, but not all, transiting
planets are indeed in the $r_{s,0} \gg r_L$ regime; ignoring magnetic
effects, Roche lobe overflow would then be a good approximation. In the
opposite limit of $\epsilon \ll 9/\lambda^2$, the solution would more
closely resemble a thermally driven wind weakly perturbed by tides.
The second assumption underlying a narrow flow through L1 is that
the mass ratio of the two bodies is near unity.  Although the tidal
expansion $r \ll D$ in eq.\ref{eq:U} ignores the difference in potential
between the L1 and L2 Lagrange points, inclusion of higher order terms
gives $U_{L2}-U_{L1} \simeq 2GM_p/(3D)$ for the potential difference
\citep{2000ssd..book.....M}.  When the ratio $2GM_p/(3Da^2)=(2/3)(\epsilon
\lambda M_p/M_\star)^{1/3} \ll 1$, the density difference between the L1
and L2 points is small, and nearly equal mass loss is expected through L1
and L2. While mass loss through L1 enters into an orbit around the star,
mass loss through L2 leads to gas in a circumbinary orbit.

MHD effects, in particular the existence of a dead zone, further limit the
applicability of the Roche lobe model. If the planet has a sufficiently
large magnetic field that the L1 Lagrange point lies inside the dead
zone, gas pressure is insufficient to open the magnetic field lines
and the flow through the L1 point is expected to be choked off. Also,
the magnetic field may torque the gas, keeping it in corotation with the
planet out to the Alfv\'en radius. By contrast, if $B^2/8\pi \ll P \simeq
\rho v^2$ at the sonic point, and $r_{s,0} \gg r_L$, magnetic stresses
and tides may be ignored the Roche-lobe model is expected to be recovered.

For the models of HD 209458b considered in this paper, inspection of Table
\ref{tab:models} shows that $r_{s,0}$, $r_L$ and the dead zone radius
$r_d$ may be within factors of a few of each other, and the situation
is more complex than the simplified Roche-lobe overflow model permits.

%%%%%%%%%%%%%%%%%%%%%%%%%%%%%%%%%%%%%%%%%%%%%%%%%%%%%%%%%%%%%%%%%%%%%

\section{Summary and Discussion} 
\label{sec:summary}

The objective of this paper was to develop a model for the upper
atmospheres of hot Jupiters, including the influence of a dynamically
important magnetic field. Our starting point (\S's \ref{sec:helmet},
\ref{sec:B}, \ref{sec:pi}, \ref{sec:dead}, \ref{sec:wind},
\ref{sec:global} and Appendix \ref{sec:mhd}) was to estimate field
strengths for hot Jupiters, and to apply the theoretical model developed
for MHD winds from stars to the case of winds escaping from the upper
atmospheres of planets. In the process, we included strong tidal
forces from the parent star (\S's \ref{sec:U}, \ref{sec:dead}, and
\ref{sec:wind}). We computed a 1D model of the temperature profile and
ionization state of the atmosphere (\S~\ref{sec:Hlayer}), and constructed
maps of neutral hydrogen column and fluid velocities
to understand the mass loss and transmission spectra of
HD 209458b (\S's \ref{sec:global}, \ref{sec:mdot}, \ref{sec:NH}, and
\ref{sec:lymanalpha}).  We contrast this model to the standard Roche-lobe
overflow model (\S~\ref{sec:roche}) and verify, {\it a posteriori},
the validity of the MHD approximation (Appendix \ref{sec:mfp}).

In section \ref{sec:B}, we discussed the application of dynamo models
to understand the magnetic field strength generated by the planet,
which is currently unconstrained by observations.  Using the recent
results of \citet{2009Natur.457..167C}, which showed that the dynamo field
increases with heat flux in the planet's core, we argued that the large
radii of hot Jupiters, and hence large core flux, imply that the magnetic
fields of inflated hot Jupiters may be larger than Jupiter's field. This
motivated exploring a wide range of possible magnetic field strengths,
both smaller and larger than Jupiter's field.

The formation of a dead zone, in which gas pressure is insufficient
to open up magnetic field lines, was motivated with a toy problem
(\S~\ref{sec:helmet}) as intuition for understanding the detailed
structure of the hydrostatic model (\S~\ref{sec:dead}). The projection
of the tidal force along magnetic field lines was used to derive the
``magnetic Roche lobe radius" (\S~\ref{sec:U}), outside of which gravity
points outward along the magnetic loop. Net gravity can point outward for
loops slightly larger than the distance to the L1-L2 Lagrange points,
even in the plane perpendicular to the star-planet line. As a result
of net outward gravity, the density may increase outward, as shown
in Figure \ref{fig:pvsr}.  We defined the key parameters $\lambda$ and
$\epsilon$, characterizing the binding energy of the gas and the strength
of tides, and their values for the observed transiting planets were given
in Figures \ref{fig:lambdaobs} and \ref{fig:epsobs}. Many close-in
planets have weakly bound atmospheres with $\lambda \la 10$, and are
subject to strong tidal forces with $\epsilon \ga 0.1$. The magnetic
field strength was characterized by the plasma $\beta$ evaluated at
the base of the atmosphere. Solutions for the radius of the dead zone
depend on the parameters $\lambda$, $\beta$ and $\epsilon$, as shown in
Figure \ref{fig:cusp}. We found that for typical parameters, the dead
zone extends to $\simeq (3-20) R$, implying that much of the volume
of the magnetosphere near the planet is occupied by bound gas with no
bulk velocity. Even gas outside the Roche-lobe radius can be static,
if the dead zone is larger than the Roche-lobe radius.

Open field lines, which are capable of supporting an outflow, were
discussed in section \ref{sec:wind}. The momentum equation along field
lines was used to compute the positions of the (slow magneto)sonic
points for a set of models using dipole geometry.  Analytic solutions
in the limit of strong and weak tides were given, which illustrated
that inward tidal forces at the magnetic poles (for a magnetic dipole
moment aligned with the orbital angular momentum axis) may eliminate the
sonic point solutions near the planet. Thus, sufficiently strong tides
effectively shut off the wind, creating a second dead zone at the poles.
%Figure \ref{fig:sonic} shows solutions for the sonic point radius for
%$\epsilon$, and Figure \ref{fig:vbase} shows the velocity at the base
%of the flow. 
Figures \ref{fig:sonic} and \ref{fig:vbase} show solutions for sonic point radius and base velocity versus footpoint position. When the sonic point position moves far from the planet,
the base velocity becomes small, and the field lines are effectively
hydrostatic. Depending on $\epsilon$ and $\beta$, the equatorial and
polar dead zones may dominate the volume near the planet. Lastly, we
estimated the asymptotic flow speed due to tides in eq.\ref{eq:vasymp},
showing that $v_{\rm asymp} \ll 100\ {\rm km\ s^{-1}}$ for the orbital
periods and stellar radii of interest. Consequently, bulk motion cannot
affect the Lyman $\alpha$ line profile at $\Delta v \ga 100\ {\rm km\
s^{-1}}$ from line center.

As a prelude to discussion of global models of the magnetosphere, and the
Lyman $\alpha$ transmission spectrum, we presented a simple spherical
model of photoionization and thermal balance (\S~\ref{sec:Hlayer}) in
order to assess the size of the ``warm" neutral H layer. We computed the
depth dependence of photoelectric heating in Figure \ref{fig:pirate},
showing that the heating drops off with pressure as a power-law, rather
than an exponential, into the atmosphere. The resulting photoelectric
heating, which we assumed was balanced by collisionally-excited Lyman
$\alpha$ cooling, gives temperatures $T \simeq (5-10) \times 10^3 \rm
K$ down to pressures $P \simeq (10-100)\ \rm nbar$.  As a result, this
neutral H layer contributes significantly to the radius, as shown in
Figure \ref{fig:pieq}. As first stressed by \citet{2010arXiv1004.1396K},
the location of the warm H layer is key in understanding the large
observed transit depths $\delta F/F \sim 5-10\%$. The transit depth due
to the layer extending upward from the H-H$^+$ ionization layer alone
is too small to explain the observations of HD 209458b, as discussed in
detail by \citet{2009ApJ...693...23M}.

Global models of the magnetosphere were constructed (\S~\ref{sec:global}),
both to compute mass loss rates (\S~\ref{sec:mdot}), and to construct
maps of the neutral hydrogen column densities for a range of parameters
as observed during transit (\S~\ref{sec:NH}).  A by-product of the warm,
deep H layer is a larger mass loss rate than in studies with more shallow H
layers \citep[e.g.,][]{2009ApJ...693...23M}. The net mass loss rates are
still insufficient to evaporate the planet, and are reduced by a factor of
3-10 due to the presence of the magnetic field for the parameters used.
The largest columns within a few $R$ of the planet occur in the dead
zones, and may receive a contribution from H atoms outside the Roche lobe,
but which are still bound to the planet.  Hence, the observation of H atoms
outside the Roche lobe alone cannot be stated as evidence for mass loss.

The 9 global models in Table \ref{tab:models} were used to compute Lyman
$\alpha$ transmission spectra in section \ref{sec:lymanalpha}. We stress
that the observational quantity most directly comparable with models of
the magnetosphere is the fractional flux decrease between in and out of
transit spectra --- this quantity is relatively independent of ISM absorption,
geocoronal contamination, and the background stellar spectrum, and is
directly computable from atmosphere models. The comparison between the
models and data for HD 209458b from \citet{2008ApJ...688.1352B} is shown
in Figure \ref{fig:absspec}. By variation of the base pressure of the warm
H layer, and temperature, models can be made to bracket the data points,
although the large error bars do not allow precise determination of the
atmosphere's parameters. Increased magnetic field is shown to increase
the transit depth, as does moving the planet further from the star.

%A comparison of the MHD wind model presented in this paper with the more
%common Roche-lobe overflow model was given in section \ref{sec:roche}.
%We showed that different regions of accretion are possible depending on
%the position of the sonic point (of an isolated body), the position of
%the L1-L2 Lagrange points, and the size of the dead zone. In particular,
%we point out that when the L1 Lagrange point occurs inside the dead zone,
%gas pressure is insufficient to open up the field lines, and a narrow
%flow through L1 is not possible. For the models of HD 209458b
%presented in this paper, our results demonstrate that mass loss from hot
%Jupiters is a more complex process than in the simplified Roche
%Lobe overflow model.  Estimates of collision rates in the atmosphere
%(Appendix B) demonstrate the validity of the MHD approximation, that
%the e-p-H gas is well-coupled collisionally at the expected pressures in
%the atmosphere, and that even neutral H gas cannot freely escape the planet.
A comparison of the MHD wind model presented in this paper with the more
commonly-used Roche-lobe overflow model was given in section \ref{sec:roche}. It was
argued that different regimes of accretion are possible depending on the
position of the sonic point (of an isolated body), the L1-L2 Lagrange
points, and the size of the dead zone. In particular, if the L1 Lagrange
point is inside the dead zone, gas pressure is insufficient to open up
the magnetic field lines, and a narrow flow through L1 is not possible. These
considerations suggest that mass loss from hot Jupiters may be more
complex than the simple Roche-lobe overflow model. Estimates of collision
rates in the atmosphere (Appendix B) demonstrate the validity of the
MHD approximation, that the e-p-H gas is well-coupled collisionally at
the expected densities and temperatures in the atmosphere, and that even
neutral H gas cannot ballistically escape the planet.

%It is therefore also important to consider the planet's magnetic field in
%our calculations, which is dynamically important for models of mass loss
%for HD 209458b.  In addition, future work that simultaneously considers
%the interaction of a hot Jupiter magnetosphere with the incoming,
%magnetized stellar wind would be useful for understanding how hot Jupiters
%interact with their parent stars.  The inclusion of additional heating
%and cooling physics in the atmosphere, as well as non-ideal MHD effects
%in deeper layers, would also help to reveal more about how observable
%quantities are coupled to structure in the magnetosphere.  More detailed
%models of these systems may also permit new methods for the discovery of exoplanets.
The model presented in this paper shows that magnetic fields may strongly
affect theoretical estimates of fluid density and velocity in the upper
atmosphere, and even the interpretation of transit depths, since neutral H
atoms outside the Roche-lobe radius may not be escaping. In future work,
we hope to include additional physical effects, such as the interaction
with the stellar wind, more detailed photoionization calculations
including heavy elements, and collisional (non-MHD) effects, which will
allow a more comprehensive physical picture of the upper atmospheres of
hot Jupiters.

\acknowledgements

We thank many colleagues for discussions, including Sean Matt on MHD wind
theory, Chris Matzner on photoionization equilibrium in dipole geometry,
Jean Michel Desert and David Sing on observations, Craig Sarazin on
cooling physics, Lars Bildsten on CV systems, and Anne Verbiscer on
the Lyman $\alpha$ scattering. We also thank the referee for feedback
that improved the paper. GT gratefully acknowledges support from NASA's
sponsorship of the Virginia Space Grant Consortium and a Teaching \&
Technology Support Partner Fellowship from the University of Virginia.
PA thanks the Kavli Institute for Theoretical Physics and participants
in the program ``Theory and Observation of Exoplanets" for a stimulating
visit. This work was supported in part by NSF (AST-0908079) and NASA
Origins (NNX10AH29G) grants. PA is an Alfred P. Sloan Fellow, and
gratefully acknowledges support from the University of Virginia's Fund
for Excellence in Science and Technology.

\appendix

\section{A. MHD wind equations }
\label{sec:mhd}

In this appendix, we present the MHD equations and discuss how currents
produced in the magnetosphere modify the field produced by the
planet's core. This discussion motivates our choice of field geometry
used in the global models.

There is a well developed literature for axisymmetric winds from
rotating, magnetized stars. An excellent review is given by
\citet{1996astro.ph..2022S}.  Here we rely heavily on the analytic
studies in \citet{1968MNRAS.138..359M} and \citet{1987MNRAS.226...57M}.  The
inclusion of the magnetic field can greatly affect the mass loss
rate and wind speed for sufficiently fast rotation. We are not aware
of detailed studies of wind launching from rotating magnetized
bodies including the non-axisymmetric tidal acceleration. We postpone
a numerical study of such a problem to a future investigation, here
using a semi-analytic treatment.

The three-dimensional MHD equations for steady isothermal flow in the frame corotating
with the planet are mass continuity
\be
\grad \cdot \left( \rho \vec{v} \right) & =&  0,
\label{eq:cont}
\ee
the Euler equation
\be
 \vec{v} \cdot \grad \vec{v} + 2\vec{\Omega}\times \vec{v}
& =&   - a^2 \grad \ln \rho - \grad U + \frac{\vec{J} \times \vec{B}}{c \rho},
\label{eq:mom}
\ee
Ohm's law for infinite conductivity
\be
\vec{E} & =& - \vec{v} \times \vec{B}/c,
\ee
the induction equation
\be
\grad \times \vec{E} & = & - \frac{1}{c} \grad \times \left( \vec{v} \times \vec{B} \right)  =   0,
\label{eq:induct}
\ee
Ampere's equation
\be
\grad \times \vec{B} & = & \frac{4\pi}{c} \vec{J},
\label{eq:ampere}
\ee
the isothermal equation of state
\be
P & =& \rho a^2,
\label{eq:energy}
\ee
and the no monopoles condition
\be
&& \grad \cdot \vec{B} = 0.
\label{eq:divB}
\ee
We have used constant $a^2$ to rewrite the
pressure gradient as $-\grad p/\rho = - a^2 \grad \ln \rho$.  The
isothermal approximation is justified in section \ref{sec:Hlayer}.
The Coriolis and centrifugal forces appear in eq.\ref{eq:mom} as
we work in a corotating frame (see section \ref{sec:U}).

To gain further insight, we rewrite eq.\ref{eq:mom}
using the vector identity $\vec{v} \cdot \grad \vec{v} = \grad (v^2/2) - \vec{v} \times
(\grad \times \vec{v})$ to obtain
\be
\grad W &=  & \vec{v} \times \left( 2\vec{\Omega} + \grad \times \vec{v} \right)
+ \frac{1}{\rho c} \vec{J} \times \vec{B}
\label{eq:mom2}
\ee
where
\be
W & \equiv & \frac{1}{2} v^2 + a^2 \ln \rho + U.
\label{eq:bernoulli}
\ee
Constants of the motion can be derived by dotting eq.\ref{eq:mom2}
with $\vec{B}$ to eliminate the Lorentz force. We find
\be
\vec{B} \cdot \grad W
& = &
 - \left( 2\vec{\Omega} + \grad \times \vec{v} \right) \cdot \left( \vec{v} \times \vec{B}
\right) = 0,
\label{eq:momdotB}
\ee
since the electric field vanishes in the co-rotating frame \citep{1996astro.ph..2022S}.
Hence $W$, the Bernoulli constant, is constant along field lines.
Another way to understand the work done on the gas
is to dot eq.\ref{eq:mom2} with $\vec{v}$:
\be
\rho \vec{v} \cdot \grad W & = & \grad \cdot \left( \rho \vec{v} W \right)
= - \frac{1}{c} \vec{v} \cdot \left( \vec{B} \times \vec{J} \right)
= \vec{J} \cdot \vec{E} = 0.
\ee
In the rotating frame, work is done on the gas by $-\grad U$,
while in the inertial frame the electromagnetic field
performs $\vec{J} \cdot \vec{E}$ work on the gas \citep{1996astro.ph..2022S}.

To understand the magnetic field structure in more detail, we take
the cross product of eq.\ref{eq:mom2} with $\vec{B}$ to obtain the equation
of trans-field force balance. Solving this equation for the component of current
perpendicular to $\vec{B}$ we find
\be
\frac{4\pi}{c} \vec{J}_\perp & \equiv & \frac{4\pi}{c} \left( \vec{J}
- \vec{b} \vec{b} \cdot \vec{J} \right)
 =
\frac{1}{v_A^2} \vec{B}
 \times \left[ \grad W + \left( 2\vec{\Omega} + \grad \times \vec{v}  \right) \times \vec{v}
\right].
\label{eq:Jperp}
\ee
Here $v_A=B/\sqrt{4\pi \rho}$ is the Alfv\'en speed.  This equation
describes the perpendicular currents that must flow in order to
achieve perpendicular force balance. In axisymmetry, this equation
is often called the modified Grad-Shafronov equation
\citep{1978JGR....83.2457H,1986ApJS...62....1L}. Perpendicular
currents arise due to either vorticity in the flow, or variation
of the Bernoulli constant from one field line to the next. In the dead zone,
the fact that $\vec{v}=0$, and the further assumption that $W$ is constant
at the base, implies that $\vec{J}_\perp=0$ in the dead zone. Parallel
currents are determined from $\vec{J}_\perp$ by charge conservation, $\grad \cdot
\vec{J}=0$.

An order of magnitude estimate for the fields $ \delta \vec{B}$ created by 
volume currents $\vec{J}_\perp$, 
compared to the planetary dynamo-generated fields $\vec{B}_p$ is
\be
\frac{ B_\perp}{B_p} & \sim & \frac{r}{B} \frac{4\pi}{c} J_\perp
\sim \frac{ {\rm max}(a^2,\ v^2,\ \Omega rv,\ (\Omega r)^2 ) }{v_{A,p}^2},
\label{eq:Bperp}
\ee
where $v_{A,p}=B_p/\sqrt{4\pi \rho}$.
The terms separated by commas on the right hand side of eq.\ref{eq:Bperp}
are estimates of the individual terms in eq.\ref{eq:Jperp}.  This estimate
shows that volume currents can only significantly perturb the field
out near the Alfv\'en radius where $v \sim \Omega r \sim v_{A,p}$.  As we now
discuss, at much smaller radii, of order the dead zone radius, the field
is already strongly perturbed by current sheets.

\citet{1968MNRAS.138..359M} and \citet{1987MNRAS.226...57M}
discussed the matching
conditions between the dead and wind zones. The finite velocity in
the wind zone acts to decrease the pressure there relative to the dead
zone. Integrating the momentum equation across the dead zone-wind zone
boundary, the total gas plus magnetic pressure must be continuous, so
that the magnetic field strength must increase moving from the dead to
the wind zone. This implies the existence of a current sheet separating
the dead and wind zone boundaries, as shown in Figure \ref{fig:cartoon}.
Letting the subscripts ``d" and ``w" denote quantities just inside the 
dead and wind zones, respectively, total pressure continuity can be
written
\be
P_d + \frac{B_d^2}{8\pi} & = & P_w + \frac{B_w^2}{8\pi}.
\label{eq:balance}
\ee
For identical conditions at the base, the Bernoulli equation relates
the pressures as $P_w \simeq P_d \exp(-v^2/2a^2)$. Considering only
the dipole field from the planet, $B_p$, and the field $\delta B
\simeq 2\pi K/c$ produced by current per unit length, $K$, 
the fields in the dead and wind zones are $B_w=B_p + \delta B$
and $B_d=B_p-\delta B$. Plugging in to eq.\ref{eq:balance} the solution
for the line density is
\be
\frac{K}{c} & = & \frac{P_d}{B_p} \left( 1 - e^{-v^2/2a^2} \right).
\ee
The ratio of the field produced by the sheet current compared to that
from the planet's core is then
\be
\frac{\delta B}{B_p} & \simeq & \frac{1}{4} \beta_d \left( 1 - e^{-v^2/2a^2} \right)
\ee
where $\beta_d=8\pi P_d/B_p^2$ is the beta for the planetary field just
inside the dead zone.  Inside the sonic point in the wind zone, $v \ll
a$ and sheet currents only slightly perturb the field since $\delta B/B
\sim \beta_d v^2/8a^2 \sim (v/2v_A)^2 \ll 1$.  Outside the sonic point,
where $v/a \gg 1$, we find $\delta B/B_p \sim \beta_d/4$, which increases
outward. Hence the field configuration is expected to be significantly
altered from the dipole outside the $\beta_d \sim 1$ point in the wind
zone. As we have assumed that $\beta \gg 1$ at the sonic point, we expect
the field to be altered in between the sonic and Alfv\'en points.

In addition to the sheet currents at the dead zone-wind zone boundary,
there is a sheet current at the equator in the wind zone.  This sheet
current causes the reversal in sign of the field near the equator,
approaching the split monopole form $\vec{B} \propto r^{-2} \vec{e}_r$ 
sufficiently distant from other current sources near the planet.
Since $K \propto B_r \propto 1/r^2$ in the wind zone, and $K\propto \beta_d$
increases in the dead zone, we expect the maximum current to occur near
the cusp in the magnetic field. The polar dead zone is expected to have
a smooth transition from dead to wind zone, as Figure \ref{fig:vbase}
shows a gradual transition. We expect volume currents in this transition,
rather than true sheet currents.

Based on these analytic estimates, an approximate field geometry
in the wind zone is roughly dipolar inside the dead zone radius
and roughly straight field lines outside.  To go beyond this
would require a detailed solution of the trans-field force balance
for the field geometry, which is beyond the scope of the present
work. 

\section{ B. mean free paths, ion-neutral drift and Ohm's law }
\label{sec:mfp}

In this section we discuss the relative motion of the e-p-H gas
as well as the magnetic field for the conditions relevant to hot Jupiters
(see Figure \ref{fig:pieq}). Equations and collision rates are taken
from \citet{2004iono.book.....S}, SN hereafter.

To simplify the calculation, we assume all three species are isothermal
with temperature $T$, and we work in the ``diffusion approximation"
in which inertial terms are ignored in the fluid equations of each
species. Let $\vec{v}_j$ be the mean velocity of species $j$, $\vec{E}$
the electric field, and $\nu_{jk}$ the momentum-transfer collision rate
between species $j$ and $k$. Momentum conservation implies $n_j m_j
\nu_{jk} = n_k m_k \nu_{kj}$. We follow \citet{1965RvPP....1..205B}
and ignore anisotropy in the collision frequencies, using the parallel
value here for simplicity. The effective gravity be denoted $\vec{g}
= - \grad U$ and the pressures are $P_j=n_jk_b T$.
The momentum equations for e, p and H are, respectively,
\be
&& -e n_e \left( \vec{E} + \frac{1}{c} \vec{v}_e \times \vec{B} \right)
- \grad P_e
+ n_e m_e \left[ \vec{g}
+ \nu_{ep} \left( \vec{v}_p - \vec{v}_e \right)
+ \nu_{eH} \left( \vec{v}_H - \vec{v}_e \right) \right]  =  0
\label{eq:emom}
\\
&& e n_p \left( \vec{E} + \frac{1}{c} \vec{v}_p \times \vec{B} \right)
- \grad P_p
+ n_p m_p \left[ \vec{g}
+ \nu_{pe} \left( \vec{v}_e - \vec{v}_p \right)
+ \nu_{pH} \left( \vec{v}_H - \vec{v}_p \right) \right]  =   0
\label{eq:pmom}
\\
&&  - \grad P_H
+ n_H m_p \left[ \vec{g}
+ \nu_{He} \left( \vec{v}_e - \vec{v}_H \right)
+ \nu_{Hp} \left( \vec{v}_p - \vec{v}_H \right) \right] = 0.
\label{eq:Hmom}
\ee
In order, the terms in eq.\ref{eq:emom} are the Lorentz force, the pressure gradient,
gravitational force, and collision drag force between e-p and e-H. We
further impose charge neutrality
\be
n_e & = &  n_p,
\label{eq:neutrality}
\ee
and define the center of mass velocity (used throughout the paper)
\be
\vec{v} & = & \frac{n_e m_e \vec{v}_e + n_p m_p \vec{v}_p + n_H m_p \vec{v}_H}
{m_e n_e + m_p n_p + m_p n_H}
\simeq 
\frac{ n_p \vec{v}_p + n_H \vec{v}_H}
{ n_p + n_H},
\label{eq:com}
\ee
where the second equality is valid in the $m_e/m_p \ll 1$ limit.

\begin{figure}[t]
\epsscale{0.5}
\plotone{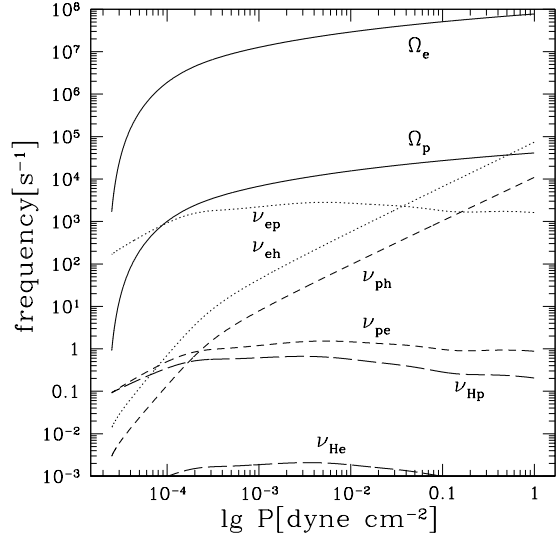}
\caption{ Gyration and collision frequencies versus depth for the model
shown in Figure \ref{fig:pieq}.
}
\label{fig:freq}
\end{figure}

\begin{deluxetable}{lcc}
\tablecolumns{3}
\tablewidth{0pc}
\tablecaption{ Collision and gyration frequencies \label{tab:freq} }
\tablehead{}
\colhead{}
\startdata
$\rm \Omega_e = \frac{eB}{m_e c} = 1.8 \times 10^7\ rad\ s^{-1} \left( B/1G \right) $ & electron cyclotron &  \\
$\rm \Omega_i = \frac{eB}{m_p c} = 9.6 \times 10^3\ rad\ s^{-1} \left( B/1G \right) $ & proton cyclotron &  \\
$\rm \nu_{eH} = 6.4\ s^{-1} \left( n_H/10^8\ cm^{-3} \right)$ & polarization & SN eq.4.88 and table 4.1  \\
$\rm \nu_{He} = 3.5\ \times 10^{-3} s^{-1} \left( n_e/10^8\ cm^{-3} \right)$ & polarization & SN eq.4.88 and table 4.1   \\
%$\rm \nu_{pp} = 60.0\ s^{-1} \left( n_p/10^8\ cm^{-3} \right) \left( 10^4\ K/T \right)^{3/2} \left( \ln \Lambda_{pp}/10 \right) $ & Coulomb & SN eq.4.140, 4.56  \\
%$\rm \nu_{ee} = 2600.0\ s^{-1} \left( n_e/10^8\ cm^{-3} \right) \left( 10^4\ K/T \right)^{3/2} \left( \ln \Lambda_{ee}/10 \right) $ & Coulomb & SN eq.4.140, 4.56  \\
$\rm \nu_{ep} = 3700.0\ s^{-1} \left( n_p/10^8\ cm^{-3} \right) \left( 10^4\ K/T \right)^{3/2} \left( \ln \Lambda_{ep}/10 \right) $ & Coulomb & SN eq.4.140, 4.56  \\
$\rm \nu_{pe} = 2.0\ s^{-1} \left( n_e/10^8\ cm^{-3} \right) \left( 10^4\ K/T \right)^{3/2} \left( \ln \Lambda_{pe}/10 \right) $ & Coulomb & SN eq.4.140, 4.56  \\
$\rm \nu_{pH} = 1.2\ s^{-1} \left( n_H/10^8\ cm^{-3} \right) \left( T/10^4\ K \right)^{1/2} $ & Charge exchange & SN table 4.5  \\
$\rm \nu_{Hp} = 1.2\ s^{-1} \left( n_p/10^8\ cm^{-3} \right) \left( T/10^4\ K \right)^{1/2} $ & Charge exchange & SN table 4.5  \\
\enddata
\end{deluxetable}

The momentum transfer and cyclotron frequencies are given in
Table \ref{tab:freq}.  They are shown as a function of depth in
Figure \ref{fig:freq} using values of $n_p$, $n_H$, $T$ and $B=B_{\rm
J,eq}(R/r)^3$ for the hydrostatic model of the equatorial dead zone shown
in Figure \ref{fig:pieq}. For these parameters, the gyration frequencies
are larger than the e and p collision frequencies over the entire H$^+$
and H layers, implying motion of both e and p perpendicular to magnetic
field lines is greatly restricted by the magnetic field. Collisions with 
e are dominated by p well into the H layer, while H dominates collisions
with p deeper than the H-H$^+$ transition. H atom collisions with p
dominate over those from e.

For a hydrogen atom traveling at a typical speed $c_H \simeq 10\
{\rm km\ s^{-1}} (T/10^4\ {\rm K})^{1/2}$, the mean free path against
collisions with p is $\simeq c_H/\nu_{Hp} \simeq 10\ {\rm km}\
(10^8\ {\rm cm^{-3}}/ n_p)$. The proton density is sufficiently large
that the mean free path is smaller than the scale height, $\simeq
r^2/\lambda R$, over the entire range shown in Figure \ref{fig:pieq}.
We conclude that, due to proximity to the star, the high temperature
and large scale height cause the density to be large enough that a fluid
treatment is appropriate. The hot Jupiter magnetospheres discussed here are
collisional, and the exobase is sufficiently distant from the planet to
be of little practical importance.  A corollary is that H atoms do not
fly ballistically through the magnetosphere, and hence acceleration by
stellar tidal gravity or radiation pressure does not cause acceleration
of H atoms away from the planet (Lyman $\alpha$ radiation pressure
is only effective in a thin outer skin where Lyman $\alpha$ optical
depth is less than unity \citep{2009ApJ...693...23M}). Rather,
acceleration induces a drift velocity, which we now estimate.

Ignoring the
$\nu_{He}$ term in eq.\ref{eq:Hmom}, the ion-neutral drift velocity is
\be
\vec{v}_H-\vec{v}_p & \simeq & \frac{1}{\nu_{Hp}}
\left( \vec{g} - \frac{1}{n_H m_p} \grad P_H \right).
\label{eq:vH}
\ee
For a simple estimate of the drift speed, we ignore the pressure
gradient term, and use fiducial values $g \simeq 10^3\ {\rm cm\ s^{-2}}$
and $\nu_{Hp} \simeq 1\ {\rm s^{-1}}$, giving $v_H-v_p \sim 10\ {\rm m\
s^{-1}}$, and a drift time over a distance $R_J$ of months.  However,
ignoring the pressure gradient is a poor approximation. In the H layer,
hydrogen atoms provide the pressure support and so hydrostatic balance
implies the quantity in parenthesis in eq.\ref{eq:vH} is small.  In the
H$^+$ layer, {\it in photoionization equilibrium}, the same cancellation
occurs, but for a different reason.  There, both protons and electrons
provide the pressure support, and so the proton scale height is $\simeq
2k_b T/m_p g$. But in photoionization equilibrium, eq.\ref{eq:ioneq}
implies $n_H \propto n_p^2$, giving hydrogen scale height $\simeq k_b
T/m_p g$, so that the terms in parenthesis in eq.\ref{eq:vH} very nearly
cancel. The deviations from photoionization equilibrium implies the
drift velocity is proportional to a factor $n_p/n_{\rm eq}$ in the H$^+$
layer and $n_{\rm eq}/n_H$ in the H layer, and the drift velocity is much
smaller than the naive estimate $\simeq g/\nu_{\rm Hp}$, except near
the H-H$^+$ transition.  Hence the drift time over a distance $\simeq
R$ is much longer than the photoionization time of $\simeq $ hrs, hence
photoionization equilibrium is a good approximation, as little diffusion
can occur in between photoionization events.

Next we discuss deviations from perfect flux freezing.
To derive Ohm's law, we follow \citet{1965RvPP....1..205B} and
solve eq.\ref{eq:Hmom} for $v_H$, plug
the result into eq.\ref{eq:emom}, and change references frames
from $\vec{v}_e$ to $\vec{v}$ in the Lorentz force, with the result
\be
\vec{E} + \frac{1}{c} \vec{v} \times \vec{B} & = & 
\left( \frac{\vec{J} \times \vec{B}}{n_e e c} \right)
\left[ \frac{\rho_p}{\rho} - \frac{\rho_H}{\rho}\frac{\nu_{He}}{\nu_H}
\right]
+ \frac{\rho_H}{\rho} \frac{1}{\nu_H c} \left( \vec{g}
- \frac{1}{\rho_H} \grad P_H \right) \times \vec{B}
+ \frac{\vec{J}}{\sigma} + \frac{m_e}{e} \vec{g} \left( 1 + \frac{\nu_{eH}}{\nu_H}
\right) - \frac{\grad P_e}{e n_e} - \frac{\nu_{eH}}{\nu_H} \frac{\grad P_H}{e n_e}.
\label{eq:ohm}
\ee
Here $\nu_H=\nu_{He}+\nu_{Hp}$, $\sigma^{-1} = (m_e/n_e e^2)(\nu_{ep}
+ \nu_{eH}\nu_{Hp}/\nu_H)$ is the conductivity, $\rho_p=m_p n_p$,
$\rho_H=m_p n_H$, and $\rho \simeq \rho_p + \rho_H$. The second term
on the left hand side is due to induction. The terms on the right hand side
are the Hall term, drift due to net force on the neutrals, the Ohmic term,
the (small) term due to gravity on the electrons, the electron pressure
gradient term, which gives rise to the charge separation field, and its
correction due to collisions with neutrals. The second term on the right
hand side may be put in the form of ``ambipolar diffusion", in astrophysical
parlance, by using the total momentum equation
\be
\rho_H \vec{g} - \grad P_H & \simeq & - \rho_p \vec{g} + \grad (P_e+P_p)
- \frac{1}{c} \vec{J} \times \vec{B},
\ee
yielding a term 
\be
 \frac{ \vec{B} \times \left( \vec{J} \times \vec{B} \right)}{\rho c^2
  \nu_H}
\ee
on the right hand side.

Applying eq.\ref{eq:ohm} to compute magnetic field evolution requires
knowledge currents and particle densities. As
argued in Appendix \ref{sec:mhd}, the cross-field currents are zero
in the dead zone if the Bernoulli constant is uniform at the base of
the atmosphere. While true for the simple case considered in this paper
(isothermal, hydrostatic equilibrium), non-isothermal conditions and/or
fluid motion at the base may induce perpendicular currents. 
%At the dead
%zone-wind zone boundary, sheet currents arise due to the gas pressure
%discontinuity. In the wind zone, volume currents are associated with the
%change in field configuration from dipole to nearly-straight field line.
%By charge conservation, parallel currents may then also exist. In the
%wind zone, currents arise in the transition from dipole geometry near
%the planet to nearly straight field lines.

We now discuss the relative size of terms in Ohm's law. In the
H$^+$ layer, the Ohmic diffusivity is $\eta=c^2/4\pi \sigma \simeq
(c^2/4\pi)(m_e \nu_{ep}/n_e e^2) \simeq 10^7\ {\rm cm^2\ s^{-1}}\ (T/10^4\
{\rm K})^{3/2}$, independent of density. Assuming Ohmic decay is balanced
through the induction term, and that the currents are of order $J \sim
(c/4\pi)(B/r)$, the required (center of mass) fluid velocity is $v
\sim \eta/r \sim 10^{-3}\ {\rm cm\ s^{-1}}$, many orders of magnitude
smaller than any characteristic velocity in the problem. Deep in
the H layer, Figure \ref{fig:freq} shows that the collision rates,
and hence diffusivity, may increase by an order of magnitude. For
nonzero cross field currents, the ratio of the Hall to the Ohmic term
is roughly $\sim \Omega_e/\nu_{e} \sim 10^4$, where $\nu_e \equiv \nu_{ep}
+ \nu_{eH}\nu_{Hp}/\nu_H$. When significant cross field currents exist, 
the Hall drift speed can be much larger than
the Ohmic drift speed, but still much smaller than the gas sound speed.
Lastly, if the
neutral drift speed has a cross field component, the second term
on the right hand side may generate a fluid velocity $v \sim (\rho_H/\rho)(v_H-v_p)$.
This drift speed is much larger than the Ohmic drift speed, although it is 
still much smaller than the sound speed. 

We conclude that, for the ionization models discussed in this paper,
the ion-neutral drift velocity and deviations from flux freezing in the
H and H$^+$ layers are small, and single-fluid MHD is a good approximation.

\end{document}